\documentclass[10pt,journal,compsoc]{IEEEtran}
\usepackage[T1]{fontenc}
\usepackage[table,xcdraw]{xcolor}
\usepackage{soul,color}
\usepackage{booktabs} % For formal tables
\usepackage{graphicx}
\usepackage{caption}	
\usepackage{subcaption}
\usepackage{todonotes}
\usepackage{url}
\usepackage{mathtools}
\usepackage{calc}
\usepackage{xparse}
\usepackage{epstopdf}
\usepackage{algorithm}
\usepackage{algpseudocode}
\usepackage{mathrsfs}
\usepackage{mathtools, cuted}

\usepackage{enumitem}
\usepackage{amsthm}
\usepackage{epstopdf}
\usepackage{multirow}
\usepackage{array}
\usepackage{ragged2e}
\newcolumntype{M}[1]{>{\centering\arraybackslash}m{#1}}

\newtheorem{definition}{Definition}

\ifCLASSOPTIONcompsoc
  \usepackage[nocompress]{cite}
\else
  \usepackage{cite}
\fi

\ifCLASSINFOpdf
\else
\fi

\newcounter{AFNumberOfComments}
\stepcounter{AFNumberOfComments}
\newcounter{MDNumberOfComments}
\stepcounter{MDNumberOfComments}
\newcounter{MHNumberOfComments}
\stepcounter{MHNumberOfComments}

\begin{document}

\title{DroneCells: Improving 5G Spectral Efficiency using Drone-mounted Flying Base Stations}

\author{Azade~Fotouhi,
        Ming~Ding,
        and~Mahbub~Hassan,% <-this % stops a space
\IEEEcompsocitemizethanks{\IEEEcompsocthanksitem A. Fotouhi and M. Hassan are with School of Computer Science and Engineering, University of New South Wales (UNSW), Sydney, Australia.\protect\\
E-mail: \{a.fotouhi,mahbub.hassan\}@unsw.edu.au
\IEEEcompsocthanksitem M. Ding is with Data61, CSIRO, Australia.\protect\\
 E-mail: ming.ding@data61.csiro.au}% <-this % stops an unwanted space
}

\IEEEtitleabstractindextext{%
	\justify
\begin{abstract}

We study a cellular networking scenario, 
called DroneCells, 
where miniaturized base stations (BSs) are mounted on flying drones to serve mobile users. 
We propose that the drones never stop, 
and move continuously within the cell in a way that reduces the distance between the BS and the serving users, 
thus potentially improving the spectral efficiency of the network. 
By considering the practical mobility constraints of commercial drones, 
we design drone mobility algorithms to improve the spectral efficiency of DroneCells. 
As the optimal problem is NP-hard, 
we propose a range of practically realizable heuristics with varying complexity and performance. 
Simulations show that, 
using the existing consumer drones, 
the proposed algorithms can readily improve spectral efficiency by 34\% and the 5-percentile packet throughput by 50\% compared to the scenario, 
where drones hover over fixed locations. 
More significant gains can be expected with more agile drones in the future.
A surprising outcome is that the drones need to fly only at minimal speeds to achieve these gains, avoiding any negative effect on drone battery lifetime. 
We further demonstrate that the optimal solution provides only modest improvements over the best heuristic algorithm, 
which employs Game Theory to make mobility decisions for drone BSs.

\end{abstract}

}

\maketitle

\IEEEdisplaynontitleabstractindextext

% \fi
\IEEEpeerreviewmaketitle

\IEEEraisesectionheading{\section{Introduction}\label{sec:introduction}}

%\IEEEPARstart{D}{rones} 
Drones are unmanned aerial vehicles flown via either remote control or autonomously using embedded mobility control software and sensors.
Historically, drones had been used mainly in military for reconnaissance purposes,
but with recent developments in light-weight battery-powered drones,
many civilian applications are emerging. One of the most important applications is to augment the coverage of the mobile communications networks.
In more detail, if base stations could be miniaturized to fit in the drone payload, they could be flown to any hard-to-reach-areas to provide coverage, where it is difficult or costly to install conventional towers. Such drone-mounted flying base stations, referred to as drone base stations (DBSs) in this paper,
can also be used to provide replacement coverage in crisis or augment coverage and capacity in high demand areas.  In fact, given the rising site rental costs \cite{gsmaopex} for the growing number of small cell deployments, DBSs can be an attractive alternative to conventional roof or pole mounted base stations.

Although the concept of DBS is still in its infancy, 
the research interest in this future technology is growing rapidly. 
Many academic researchers are now actively working in the area \cite{Sharma2016,Takaishi2016}, 
while industry players are also beginning to join the game. 
Nokia has recently developed an ultra miniaturized 4G base station weighing only 2Kg, 
which was successfully mounted on a commercial quad-copter to provide coverage over a remote area in Scotland \cite{nokiadrone}. 
This successful demonstration proves that the underlying hardware technology for DBS has matured.

Recent studies \cite{al2014optimal,yaliniz2016efficient} on DBS mainly focused on finding the optimum location for the drones to \textit{float} or \textit{hover} so that the coverage is maximized. 
In this paper, 
we push the potential of DBS one step further. 
Specifically, 
we propose that, 
instead of hovering over a fixed location, 
a DBS should move constantly within its cell boundary to continuously reduce the distance to the active users. 
The decreasing BS-to-user distance should result in better received signal strength for all users, 
improving the overall spectral efficiency of the network and the quality of service (QoS), 
especially for the cell-edge users, 
which usually suffer from inferior performance in conventional cellular networks.
For clarity, 
we will refer to the proposed \textit{constantly moving} DBS as an agile-DBS hereafter.

Realization of agile-DBS faces significant research challenges. 
More specifically, 
drones have practical agility constraints in terms of flying speed,
turning angles, 
and the maximum frequency at which its mobility parameters can be updated. 
Besides, 
there is the issue of mechanical energy consumption, 
which must be conserved for battery-operated drones. 
Generally speaking, 
it is desirable that the continuous mobility of an agile-DBS should not drain the battery faster than the hovering DBS. 
Finally, 
autonomous mobility control of the drone would require low-complexity and thus practically realizable algorithms to move the drones in a way so that the maximum spectral efficiency can be achieved under realistic traffic scenarios. 
Given that the interference between the cells becomes more complicated when the BSs do not stay at the same location, 
it becomes more challenging to find a movement path that will minimize such interference.

To the best of our knowledge, 
the issues related to agile-DBS have not been adequately analyzed in the literature. 
The novelty and contributions of our paper can be summarized as follows. 
We propose the novel DroneCells scenario where drones \textit{constantly move} within the cell with the objective of serving the users from a closer distance and thus improve spectral efficiency of the network and the QoS of cell-edge users. 
Based on experimentally derived agility constraints, 
we propose three practically feasible drone mobility control algorithms with varying complexity and performance. 
Simulations show that, 
using the existing consumer drones, 
the proposed mobility heuristics can readily improve spectral efficiency by 34\% and the 5-percentile packet throughput by 50\% compared to the scenario where drones hover over fixed locations. As more agile drones become available, 
we can expect more significant gains in the future. 
A surprising outcome is that the drones need to fly only at minimal speeds to achieve these gains, making it possible to avoid any negative effect on drone battery lifetime. 
We further demonstrate that the optimal solution provides only modest improvements over the best heuristic algorithm, 
which employs Game Theory to make mobility decisions for drone base stations.

%  Section \ref{sec:analyseagility} and ~\ref{sec:experiments},

The rest of the paper is structured as follows. Related work is reviewed in Section~\ref{sec:relatedwork} followed by the system model of the proposed DroneCells networks in Section~\ref{sec:systemmodel}. We introduce our drone mobility algorithms in Section~\ref{sec:proposedstr}.
Section \ref{sec:evaluation} focusses on performance evaluation of the proposed algorithms. Finally, the conclusion is discussed in Section~\ref{sec:conclusion}.


\section{Related Work}\label{sec:relatedwork}

Drones have been considered recently in the context of wireless networks due to their flexibility and agility.
In this section,
we review the drone-related researches in three different categories as follows.

\subsection{Modeling}
A new form of network consisting of flying UAVs with special characteristics like mobility,
channel model,
and energy consumption is defined in \cite{Bekmezci20131254}, and named Flying Ad hoc Networks or FANET.
Due to special characteristics of UAVs and drones,
realistic path loss and fading models are needed for these networks which are the subject of studies such as \cite{al2014optimal,7194055,7037248}.
In these studies, an elevation angle dependent model is proposed for air-to-ground communication.

\subsection{Optimal Drone Deployment}
Deploying drones in the optimal locations to fulfil different network objectives are investigated in the literature as well.
Al-Hourani et al. \cite{al2014optimal} provided an analytical model to find an optimal altitude for one UAV providing the maximum coverage of the area.
A service threshold in terms of maximum allowable path loss is defined in this model.
Another recent study by Mozaffari et al. \cite{7510870} involves finding the optimal cell boundaries and deployment locations for multiple non-interfering UAVs.
The objective of this study is to minimize the total transmission power of UAVs.
Optimal placement of UAVs to deal with disaster situation and improve public safety communication is addressed in \cite{7122576}.
Brute force search is used to find the optimal location of UAVs in the target area.
Moreover,
authors in \cite{yaliniz2016efficient} discussed finding the 3D optimal location for deploying a drone cell to provide services for the maximum number of users satisfying their SNR (Signal to Noise Ratio) constraints.

\subsection{Mobile Drones}
Rather than deploying UAVs in optimal location,
dynamic movement of UAVs are also investigated in the literature.
Maintaining MANET connectivity is discussed in \cite{1683473,4787021} through controlling the movement of one single UAV.
Minimal spanning tree model is used to control the UAV movement to improve connectivity for mobile ad hoc ground users.
Improving the probability of end-to-end connection between ground users through multi-hop UAV communication is studied in \cite{Takaishi2016}.
In this work,
the centre and the radius of circular trajectory for UAVs
are adjusted for better performance.
Motion control of UAVs' chain to improve the link capacity between two mobile nodes is explored in \cite{7237535}.
%where UAVs are acting as relay chain between the nodes.
Artificial Potential Field model is used to control the speed and heading angle of UAVs.
Moreover,
\cite{7572068} studied the throughput maximization for communication between a fixed source and destination through a mobile UAV relay.
The problem of trajectory optimization given a fixed power allocation is formulated as a non-convex optimization problem.
%}

%\textcolor[rgb]{0.50,0.51,0.53}{
Adapting the location of one single UAV acting as a relay to collect data from mobile users and forward them to another base station is studied in \cite{6214709}.
In this work,
it is assumed that the UAV can predict users' location using any position prediction algorithm.
The goal is to optimize the achievable uplink rate for users.
In our previous work \cite{7848883},
we proposed dynamic mobility control for a single drone base station to maximize the spectral efficiency of download link.
It was extended to multi-cell scenario in \cite{wowmom_main_arxiv}, assuming no limitation for drones' movements.

% Another example of dynamic trajectory control for single drone base station for maximizing the spectral efficiency of download link is considered in \cite{7848883}. % where mobile users transmit data to DBS. The DBS adjusts its position constantly to provide higher spectral efficiency.
%}

Constantly moving drone cells in a network including mobile users to improve the spectral efficiency is not explored enough in the literature. In this work, the practical limitation of drones are investigated through experiments and simulations, applied in our proposed methods. Moreover, we analysed the impact of different network elements and drone hardware parameters on the achievable gain by DroneCells.

%(1) multiple mobile interfering drones (2) which are supposed to provide services for mobile users in the cellular networks. (3) Trajectory of drones is compatible with the real practical commercial drones' features and limitation, (4) and movement algorithms are simple enough to be used in real-time application. Moreover, (5) the realistic propagation model for air-to-ground communication is used in this work.

\section{System Model} \label{sec:systemmodel}

In this section, we define the various elements that make up the proposed DroneCells. In particular, we describe the network, traffic, channel, resource allocation, user mobility, and drone mobility models.

\subsection{Network Model} \label{subsec:network_model}
\begin{figure}
	\centering
	\includegraphics[scale=0.25]{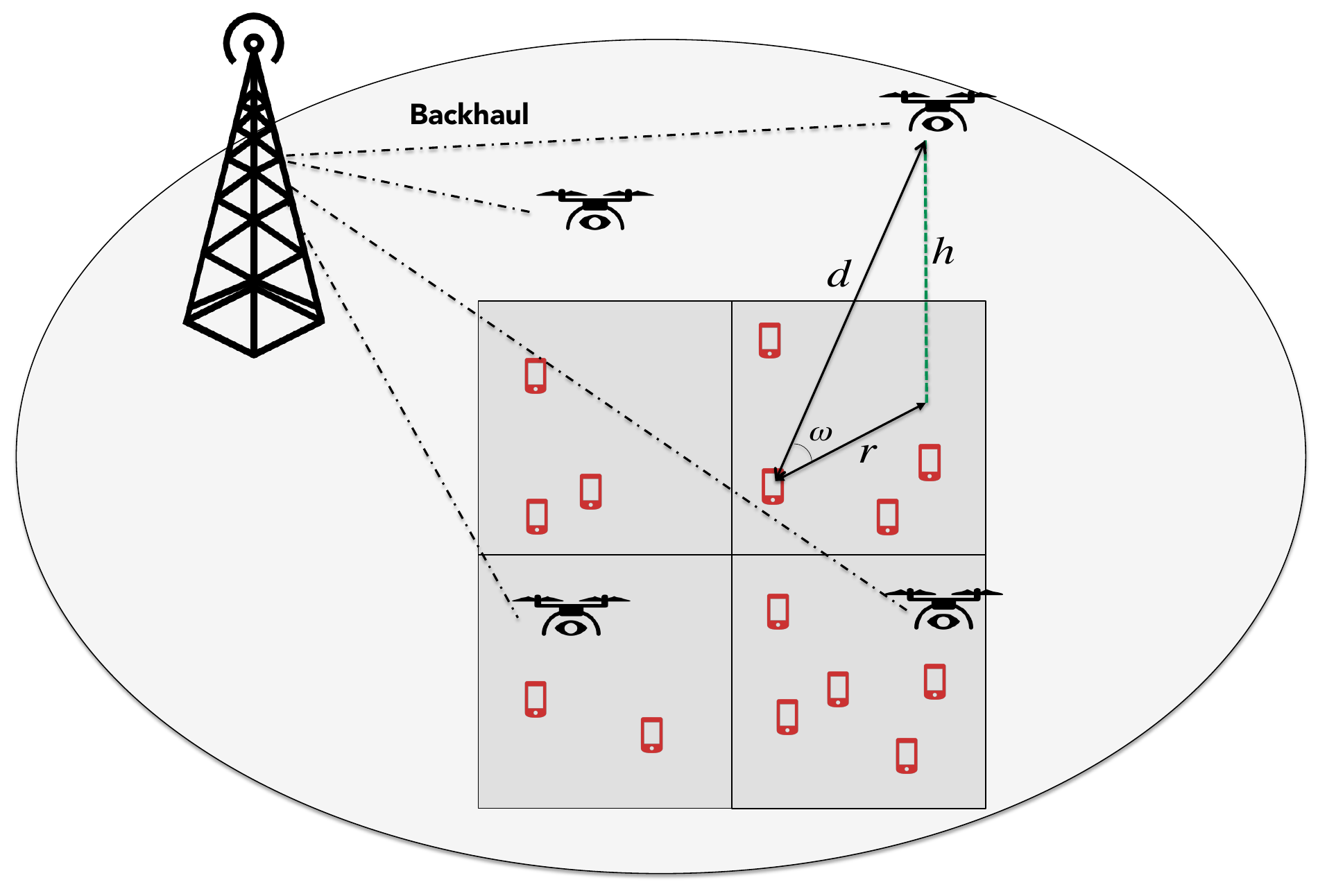}
	\caption{A 2x2 (4 cells) network model of DroneCells.}
	\label{fig:arch}
\end{figure}
Figure \ref{fig:arch} shows the DronceCells network architecture. We assume a grid of $N$ cells, where each cell is a $l (m) \times l (m)$ square served by a single DBS.
Each DBS, which may be connected to a nearby macro cell tower with a wireless backhaul link, is responsible for providing wireless communication services for the users in its cell (fronthaul). In each cell, there are $U$ mobile users associated to the DBS of that cell. The set of all drones and the set of all users are denoted by $\mathcal{U}$ and $\mathcal{N}$, respectively. Apparently, $|{\mathcal{N}}| = N$ and $|{\mathcal{U}}|= U.N$ should be satisfied, where $|.|$ outputs the size of a set.

We consider orthogonal frequency allocation between the backhaul and the fronthaul, which means that we do not have any interference between these two links. All drones, however, use the same frequency band, thus creating the potential for inter-cell interference. It is assumed that wireless transmission from a DBS can create interference on mobile users in neighbour cells up to $\kappa$ meter. The interference beyond $\kappa$ meter is negligible. We further assume that each DBS is transmitting data to users using a fixed transmission power of $p_{tx}$ (watt), total bandwidth $B$ (in Hz) with central carrier frequency of $f$ (Hz).

The \textit{ground distance} or the two-dimensional (2D) distance between user $u \in {\mathcal{U}}$ and drone $n\in {\mathcal{N}}$ is defined by the distance between the user and the projection of the drone location onto the ground, denoted by $r_{u,n}$. The \textit{euclidean distance} or the three-dimensional (3D) distance between user $u$ and drone $n$ is presented by $d_{u,n} = \sqrt{r_{u,n}^2 +h^2}$.

\subsection{Traffic Model}\label{subsec:traffic_model}

Each user follows the traffic model recommended by 3GPP \cite{3gpp36814}, which essentially dictates that a user device continuously alternates between \textit{reading time} and \textit{transmission time}:
\begin{itemize}
  \item The reading time is defined as the time interval between the end of the download of the previous data packet and the request for the next one. It follows exponential distribution with a mean of $\lambda$ (sec).
  \item The transmission time ($\tau$ sec) is defined as the time interval between the request of a data packet and the end of its download. The actual transmission time depends on the packet length and the amount of resource allocated to the user.
\end{itemize}
All packets are assumed to have a fixed size of $s$ (MByte). During a transmission time $\tau$, the associated user is called an \textit{active} user. The set of all active users in a cell $n$ at a specific time $t$ is denoted by ${\mathcal{Q}}_{n}(t)$ ($0 \leq |{\mathcal{Q}}_{n}(t)
| \leq U$).

\subsection{Channel Model} \label{subsec:channelmodel}

The wireless channel between the DBS and the mobile user on the ground is modelled following the recently introduced
%In this paper,
%we consider a practical path loss model incorporating both LoS (Line of Sight) transmissions and NLoS (Non Line of Sight) transmissions.
%More specifically,
probabilistic LoS model~\cite{al2014optimal,7194055}, in which the probability of having a LoS connection between a drone and its user depends on the \textit{elevation angle} ($\omega$) of the transmission link. According to~\cite{al2014optimal},
the LoS probability function is expressed as
\begin{equation}
P^{LoS}(u,n) = \frac{1}{1+\alpha exp(-\beta[\omega -\alpha])},
\label{eq:plos}
\end{equation}
where $\alpha$ and $\beta$ are environment-dependent constants,
$\omega$ equals to $arctan(h/r_{u,n})$ in degree,
and $h$ denotes the drone height.
As a result of (\ref{eq:plos}),
the probability of having a NLoS connection can be written as
\begin{equation}
P^{NLoS}(u,n) = 1 - P^{LoS}(u,n).
\label{eq:pnlos}
\end{equation}

From (\ref{eq:plos}) and (\ref{eq:pnlos}), the path loss in dB can be modeled as
\begin{equation}
\eta_{path}(u,n) = A_{path} + 10\gamma_{path}\log_{10}(d_{u,n}),
\label{eq:pathloss}
\end{equation}
where the string variable \textit{"path"} takes the value of {"LoS"} and {"NLoS"} for the LoS and the NLoS cases, respectively.
In addition,
$A_{path}$ is the path loss at the reference distance (1 meter) and $\gamma_{path}$ is the path loss exponent,
both obtainable from field tests \cite{TR36.828}.

\subsection{Resource Allocation} \label{subsec:resourceallocation}

Each cell has a total bandwidth of $B$ Hz, which has to be shared by all active users. Time is slotted with slot lengths of $t_{r}$ sec, and the DBS updates resource allocation every $t_{r}$ sec. We consider two different resource allocation models, \textit{equal share} and \textit{channel-quality-based (CQ-based)} allocation. The \textit{equal share} model aims to allocate resources fairly among all users, hence, in each resource allocation slot (RAS), the DBS simply divides the total bandwidth \textit{equally} among all the active users. The CQ-based allocation model, on the other hand, aims to maximize the spectral efficiency of the network at the expense of fairness. At each RAS, the CQ-based model allocates the whole bandwidth to only one active user who has the highest channel quality.

%
%In this method,
%each drone tries to serve all the active users simultaneously by equally dividing the whole available bandwidth among them.
%Despite the selected direction and location of the drone at any time,
%the drone simply divides the total bandwidth \textit{equally} among all the active users in each resource allocation time slot while moving towards the selected direction.
%When a transmission for a request finishes,
%the amount of released bandwidth is shared equally among the unfinished requests.
%Subsequently,
%in this method all active users of the cells at each time slot will receive data from their drone of interests.
%
%Note that the equal bandwidth division ensures fairness among the active users.
%Other bandwidth allocation methods can also be considered in our work,
%to achieve different levels of tradeoff between performance and fairness.
%It is straightforward to do so,
%and hence we omit such discussion for brevity.

%\subsubsection{Signal-Strength-based Allocation} \label{sec:tdma}
%In this method,
%instead of allocating bandwidth equally to all the active users,
%the whole bandwidth is devoted to only one user during each resource allocation time slot.
%Among all the active users,
%the one that has the maximum value received signal strength is selected.

\subsection{Spectral Efficiency}

The main motivation for the proposed \textit{constant movement} of the DBSs is to improve the spectral efficiency of DroneCells. In this section, we explain how spectral efficiency of DroneCells can be calculated at any given instant of time.
%%%%%%%%
For defining the spectral efficiency, we first need to define the received signal power.
The received signal power, $S^{path}(u,n)$ (watt), of an active user $u$ associated to drone $n$ can be obtained by
\begin{align}
\begin{split}
S^{path}(u,n)&=\frac{b_u}{B} \times p_{tx} \times 10^{\frac{-\eta_{path}(u,n)}{10}}\\
&=\frac{b_u}{B} \times p_{tx} \times A'_{path} \times d_{u,n}^{-\gamma_{path}},
\end{split}
\label{eq:rcvpower}
\end{align}
where $A'_{path} = 10^{\frac{-A_{path}}{10}}$, and $b_u$ is the allocated bandwidth to the user with $0 \leq b_u \leq B$.

Moreover, the total noise power, $N_u$ (watt), for an active user $u$ including the thermal noise power and the user equipment noise figure,
can be represented by \cite{thermalnoise}
\begin{equation}
N_u = 10^{ \frac{-174+\delta_{ue}}{10}}\times{b_u}\times10^{-3},
\label{eq:noise}
\end{equation}
where $\delta_{ue}$ (dB) is the user equipment noise figure.

Accordingly, the \textit{Signal to Interference plus Noise Ratio (SINR)} of user $u$ associated to drone $n$ can be expressed as
\begin{align}
\begin{split}
SINR^{path}(u,n)&=\frac{S^{path}(u,n)}{I_u+N_u}\\
&= \frac{S^{path}(u,n)}{\big(\sum_{i \in {\mathcal{N}}, i \not= n, r_{u,i} \leq \kappa } S^{path}(u,i)\big)+N_u},
\end{split}
\label{eq:sinr}
\end{align}
where $I_u$ (watt) represents the interference signal from neighbour cells received by user $u$ \footnote{Note that in this paper we focus on the analysis of small cell networks (SCNs) with an orthogonal deployment in the existing macrocell networks,
where small cells and macrocells operate on different frequency spectrum, i.e., Small Cell Scenario \#2a defined in~\cite{TR36.872}. As such, DBSs interfere only with each other, but not with the macro cells.
Indeed, the orthogonal deployment of dense SCNs within the existing macrocell networks has been selected as the workhorse for capacity enhancement in the 3rd Generation Partnership Project (3GPP) 4th-generation (4G) and the 5th-generation (5G) networks.
This is due to its large spectrum reuse and its easy management~\cite{Tutor_smallcell};
the latter one arising from its low interaction with the macrocell tier, e.g., no inter-tier interference.}.

Then, the \textit{spectral efficiency (SE)} (bps/Hz) of an active user $u$ associated with drone $n$ can be formulated according to the Shannon Capacity Theorem as~\cite{Book_Proakis}
\begin{align}
\begin{split}
\Phi^{path}(u,n) = \log_2 (1+SINR^{path}(u,n)).
\end{split}
\label{eq:individualspec}
\end{align}

Given the probabilistic channel model,
the average SE for user $u$ can be expressed as
\begin{align}
\begin{split}
\bar{\Phi}(u,n) =&P^{LoS}(u,n)  \Big( \log_2 (1+\frac{S^{LoS}(u,n)}{I_u+N_u})  \Big) \\
+& P^{NLoS}(u,n)  \Big( \log_2 (1+\frac{S^{NLoS}(u,n)}{I_u+N_u})  \Big).
\end{split}
\label{eq:averagespec}
\end{align}

Next, the average SE for a drone $n$ can be computed from
\begin{align}
\begin{split}
\bar{\Phi}(n) = \frac{\sum_{u \in {\mathcal{Q}}_{n}} \bar{\Phi}(u,n) }{|{\mathcal{Q}}_{n}|}.
\end{split}
\label{eq:secell}
\end{align}

Consequently, the average SE of the considered $N$-cell system can be obtained by
\begin{align}
\begin{split}
\bar{\Phi} = \frac{\sum_{n=1}^{N} \bar{\Phi}(n) }{N}.
\end{split}
\label{eq:sesystem}
\end{align}

%\mahbub{It is not clear why we have "path" in some of the notations.}
%
%\azade{it is described in the "Channel Model Section" that \textit{"path"} takes the value of {"LoS"} and {"NLoS"} for the LoS and the NLoS cases, respectively. Each case of  {"LoS"} and {"NLoS"} have a probability. As a result, the average value for a parameter, can be calculated by Prob\_LoS $\times$ (value of parameter having LoS)+ Prob\_NLoS $\times$ (value of parameter having NLoS) }
%%%%%%

\subsection{User Mobility}\label{subsesc:usermobility}

Users within a cell are moving according to the Random Way Point (RWP) mobility model, which is commonly used for studying
the users mobility in cellular networks \cite{rwp1,rwp2,rwp3}.
%recommended by 3GPP \cite{3gpp36814}.
In this model, each user selects a random destination within the cell independent of other users, and moves there following a straight trajectory with a constant speed selected randomly from a given range.
Upon reaching the destination, users may pause for a while before continuing to move to another destination. The pause duration is also chosen randomly from a given range.

\subsection{DBS Mobility} \label{subsec:dronemobility}

Although drones are capable of moving in 3D space, recent literature suggests \cite{7842150} that 10 meter is the optimal height for positioning a small cell base station. Lowering the antenna height below 10 m causes coverage issues, while higher than 10 m increases interference with the neighbour cells. We therefore fix the height of the DBS at 10 m, and consider DBS mobility in the 2D plane only.

In theory, many different 2D mobility models could be considered for the DBS. Some of these models may require the drone to stop at some location before changing direction and move again. Frequent stopping and starting, however, would introduce delays and consume more battery energy. For drones without rotors (drones with wings), it may be very challenging to actually stop and hover at a fixed location. For operational convenience, we therefore choose a simple mobility model that completely avoids stops and starts for the drone. Instead, the drone \textit{continues to move} in the 2D space with a constant linear speed of $v (m/s)$, but updates its decision to change direction every $t_{m}$ sec, hereafter called \textit{direction update interval}. The proposed \textit{continuously moving} model is thus applicable to all types of drones, with or without rotors.

\begin{figure}
	\centering
	\includegraphics[scale=0.35]{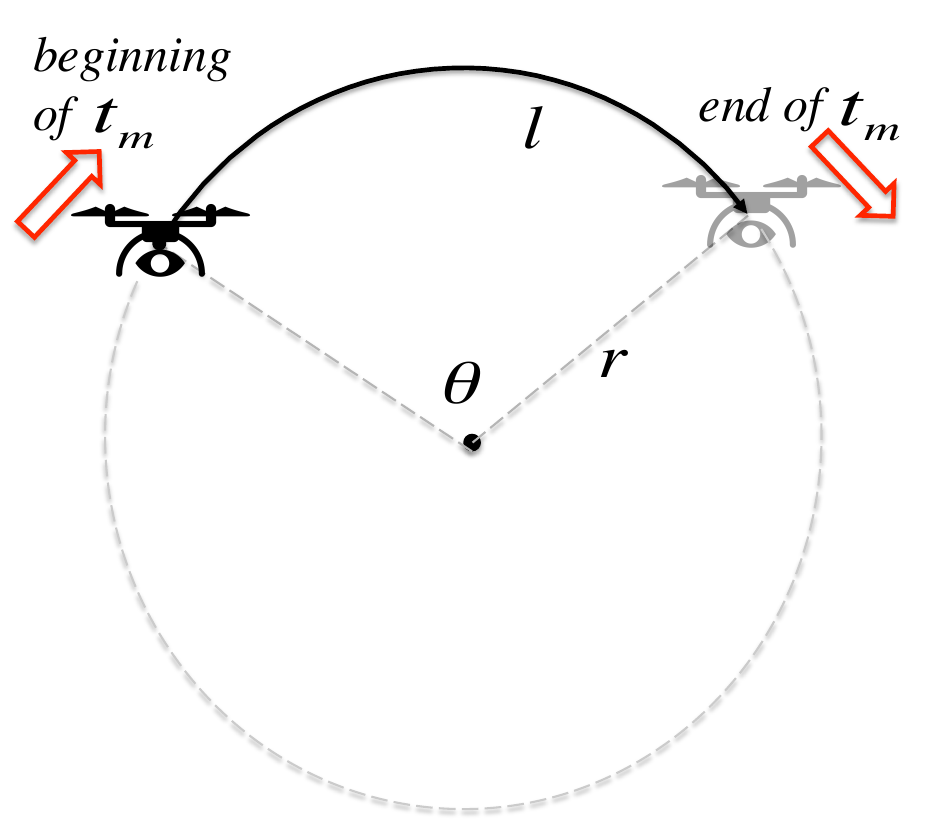}
	\caption{Drone path while taking a turn.}
	\label{fig:agilitymodel}
	%\vspace{-0.4cm}
\end{figure}

Even when the drone is turning to change the direction, it continues to travel at the constant linear speed of $v (m/s)$. As a result, during turning, the drone`s path will lie along the arc of a circle as shown in Figure \ref{fig:agilitymodel}. The radius of the circle is given by $r =\displaystyle \frac{v^2}{a}$, where $a$ is the centripetal acceleration of the drone.

Assuming a target turning angle of $\theta$ within a time interval of $t_m$, the required acceleration is obtained as \cite{Shanmugavel20101084,Agility1998}
\begin{align}
\begin{split}
a=\frac{v \times \theta}{t_m}.
\end{split}
\label{eq:agilityformula}
\end{align}

Similarly, given a maximum possible acceleration of $a_{max}$, the drone can only turn a maximum angle of $\theta_{max} = \displaystyle \frac{a_{max} \times t_m}{v}$. We use these relationships to define the 2D mobility of the DBS as follows.

At every $t_m$, the DBS chooses an angle, $\theta$, between $\pm$[0,$ \theta_{max} $]
%$\displaystyle \frac{a_{max} \times t_m}{v}$]
and starts to complete the turn at the end of next $t_m$ sec. Depending on the selected angle to turn ($\theta$), the drone would apply an acceleration of $a=\displaystyle \frac{v \times \theta}{t_m}$, which lies between [0,$a_{max}$]. Note that the range of available values for the parameters $\displaystyle v$, $\displaystyle a_{max}$, and $t_m$ are dependent on particular drone hardware, which we will investigate later in Section \ref{sec:evaluation}.

A key question is: what angle the DBS should choose to turn at the start of every $t_m$ interval? This is the subject of the DBS mobility algorithm (DMA) design, which will influence the DBS-to-user distance ($r$), the elevation angle ($\omega$), and ultimately the spectral efficiency of DroneCells. The proposed DBS mobility algorithms are explained in the following section.

\section{DBS Mobility Algorithms}\label{sec:proposedstr}
%describing abbr. HOV, SNR , and default one

DBSs need to move in a way that increases the overall spectral efficiency of the system. DBSs are constantly moving and updating their directions every $t_m$ sec. The task of DMA is to select the new direction (turning angle) at the start of every $t_m$ interval.
Several factors make the selection of the new direction a challenging problem. First, the DBS will continue to follow the path specified by the turning angle selected at the \textit{start} of the interval for the next $t_m$ seconds. This path cannot be changed in the middle of $t_m$ despite any further changes in mobile user population and traffic in the system. Second, spectral efficiency is affected not only by the DBS-to-user distances in the current cell, but also due to interference from other moving DBSs in other cells.

Given the 2D location of a drone $n$ at the start of current update interval denoted by $[x_{n}^t, y_{n}^t]$,
then by taking a turn of $\theta_{n}$ rad,
the drone will move along a candidate circle segment,
where the drone location $[x_{n}^{t'}, y_{n}^{t'}]$ on the segment at any time $t'$ during the update interval can be calculated by
\begin{align}
\begin{split}
\left[ \begin{array}{c} x_{n}^{t'} \\ y_{n}^{t'} \end{array} \right] &=
R \times \left[ \begin{array}{c} x_{n}^{t}-{cx}_{n}^t \\ y_{n}^{t}-{cy}_{n}^t \end{array} \right] +
\left[ \begin{array}{c} {cx}_{n}^t \\ {cy}_{n}^t \end{array} \right];\\
R &= \begin{bmatrix} \cos(\theta_{n}) & -\sin(\theta_{n}) \\ \sin(\theta_{n}) &\cos(\theta_{n}) \end{bmatrix},
\end{split}
\label{eq:nextdronelocation}
\end{align}
where $R$ is the rotation matrix,
and $[cx_{n}^t, cy_{n}^t]$ denotes the coordinates of the circle centre of the segment.
The drone location at $t'$ is denoted by $[x_{n}^{t'}, y_{n}^{t'}]$,
where $t'\in[t\quad t+t_m)$.
Note that $\theta_{n}$  can take any value satisfying the drone constraints.

We first study the optimal DMA that assumes knowledge of the whole system and then propose three heuristics with decreasing complexity. In all of these algorithms, the DBS chooses the direction that would bring it closer to the central point if \textit{no active users are detected}. In other words, in the absence of any activity, the drone would continue to move in the vicinity of the central point of the cell.

We have not included any explicit measures in our algorithms to provide \textit{absolute guarantees} for the DBSs not to cross the cell border, which would be too restrictive and limit the spectral efficiency gains. Instead, the proposed algorithms are designed for the DBSs to best serve the users \textit{within their respective cells}, which act as an \textit{invisible force} for the drones to stay within the cell and quickly head back to the cell if they happen to stray outside the cell boundary.  As such, DBSs moving outside the cell should be rare events, as we will demonstrate later in Section \ref{sec:evaluation} .

%\mahbub{Improve Figure \ref{fig:candidate}.}

\subsection{Optimal DMA}\label{subsec:opt_alg}

\begin{figure}
\centering
\includegraphics[scale = 0.25]{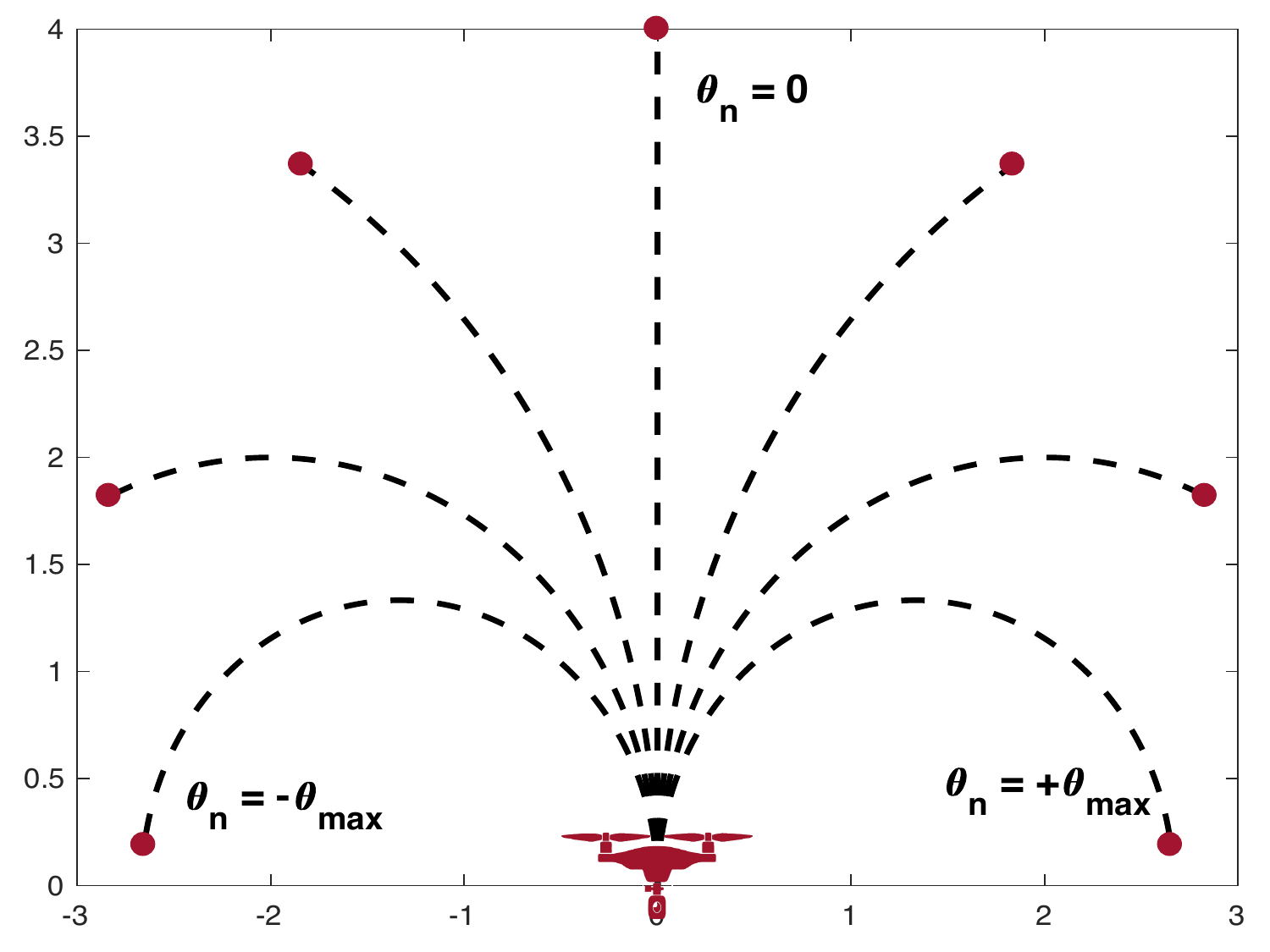}
\caption{Candidate paths for drone $n$ during a time slot}
\label{fig:candidate}
\end{figure}

Our objective is to maximize the SE of the system considering mobile drones and mobile users, 
which is non-convex problem with a feasible set of continuous directions.  
%Given the continuous feasible set of directions for drones to choose from, 
%it is a non-convex problem.
To exhaustively search for the solution, 
we discretize all turning options into a finite set of $[-\theta_{max},\dots,-2g,-g,0,g,2g,\dots,\theta_{max}]$, 
where $g=\displaystyle \frac{2\theta_{max}}{G-1}$ with $G$ representing the total number of turning options. 
Assuming that $-\theta_{max}$, $\theta_{max}$, and $0$ are the three minimum options, 
$G$ can take any odd integer values starting from 3. 
Figure \ref{fig:candidate} shows 7 candidate paths ($G$=7) for a drone for the next $t_m$ seconds.

%where $\theta_{n}$ (degree) denotes the turning angle for the drone $n$ regarding to its current moving direction.

By taking any of the possible directions/paths,
the drone can obtain a different spectral efficiency.
To calculate the spectral efficiency for a drone during a candidate path, 
for simplicity, 
we assume that all users are stayed in their initial locations at the start of update interval $t_m$. 
Moreover, 
a set of points on the taken path is selected to calculate the spectral efficiency. 
Then the spectral efficiency of a path is the average of the spectral efficiency of points on the path.

The goal of the drone mobility control is to choose a direction that maximizes the average spectral efficiency of the system.
%on the selected paths.
Hence,
the optimization problem (\textit{OPT}) at the start of each update interval can be formulated as
\begin{align}
 \begin{split}
 (\theta^*_1,\dots,\theta^*_N)  &=  \text{arg}\,\max \bar{\Phi}
 \\
 s.t. \quad &\theta^*_n  \in [-\theta_{max}^{v}:g:\theta_{max}^{v}]  \quad \forall\ n \in \mathcal{N},
\end{split}
 \label{eq:opt}
\end{align}
 %& \theta_n \in [0,2\pi)  \quad \forall\ n \in \mathcal{N}
where $ (\theta^*_n)$ denotes the optimal direction selected by drone $n$ at the start of update interval,
and $\theta_{max}^{v}$ is the maximum feasible turning angle for the drone flying with the speed $v$.

Each drone can choose its direction from $G$ candidate ones.
Therefore,
solving the optimization problem requires searching over $G^N$ cases to find the optimal direction for all $N$ drones,
which is an NP-hard problem.
Thus,
it is impractical to find the optimal solution for a large number of drones.
Here,
we propose heuristic strategies described in the next section to find the moving directions for the drones.
To verify the accuracy of the solutions from heuristic strategies,
we have also investigated the optimal solution for small number of drones and compared them with the proposed strategies.

\subsection{Game Theory DMA}\label{subsec:gt_alg}
In this section, we apply game theory to solve the direction selection problem for the interfering DBSs with less complexity than the optimal DMA discussed in the previous section. In particular, we formulate the direction selection as a non-cooperative game played by all the interfering DBSs in the system. The game is played at the start of each $t_m$ interval and the decisions leading to the Nash equilibrium are used by the DBSs to update their directions. Hereafter, we called this algorithm \textit{GT}.

The strategic game is described by ${\mathcal G} = ({\mathcal P},\{{\mathcal A}_p\},u_p)$ where
${\mathcal P}$ is the set of finite players (DBSs), ${\mathcal A}_p $ is the set of actions ($G$ turning angles) for each player, and $u_p$ is the utility function of each player. In this paper, ${\mathcal P} = \{ 1,2,\dots, P\}$ is the set of DBSs with at least one active user in their cell. The set of actions for a DBS/player $p$ can be expressed as
\begin{align}
 \begin{split}
 {\mathcal A}_p  = \{\theta_{p}:\theta_{p} \in [-\theta_{max}^v,\dots,-2g,-g,0,g,2g,\dots,\theta_{max}^v)]\}
\end{split}
 \label{eq:action_n}
\end{align}
where $g$ is the turning angle step. Moreover, we define action space ${\mathcal A} = {\mathcal A}_1 \times {\mathcal A}_2 \times \dots \times {\mathcal A}_P $, as the Cartesian product of the set of actions of all players.

%According to the definition of actions for each drone, here we have a finite game.

$u_p:{\mathcal A}\rightarrow {\rm I\!R}  $ is the utility function for each player $p$, that maps any member of the action space, $\theta \in {\mathcal A}$, to a numerical real number. We may denote the utility function of each player as $u_p(\theta_p, \theta_{-p})$, where $\theta_{-p}$ presents the action of all players except $p$. The utility function for each player is defined by the spectral efficiency of that player given the action of all players, as follows

\begin{align}
 \begin{split}
u_p(\theta) =	 u_p(\theta_p, \theta_{-p}) =  \bar{ {\Phi}}(p),
\end{split}
 \label{eq:utilityfunc_def}
\end{align}

In non-cooperative game, each player independently tries to find an action that maximizes its own utility, however its decision is influenced by the action of other players:

\begin{align}
 \begin{split}
 \text{arg}\,\max\limits_{\forall  {\theta_{p}\in {\mathcal A}_{p}}}\ u_p(\theta_p, \theta_{-p})  \quad \forall p \in  \mathcal{P}
\end{split}
 \label{eq:maxutfun_game}
\end{align}

A pure Nash Equilibrium is a convergence point where no player has an incentive to deviate from it by changing its action, defined as:

\begin{definition}
A member of action space, $\theta = (\theta^*_1,\theta^*_2,\dots,\theta^*_P)$, is a pure Nash Equilibrium (NE) if and only if
\begin{align}
 \begin{split}
u_p(\theta_p^*, \theta_{-p}^*) \geq u_p(\theta_p, \theta_{-p}^*) \quad \forall \theta_p\in {\mathcal A}_{p} \ {\rm and} \ \forall p \in  \mathcal{P}
\end{split}
 \label{eq:NE_def}
\end{align}
\end{definition}

The proposed utility function requires each player be aware of the action selected by other players. %{\color{red}We considered a signaling system where each drone base station can communicate with other interfering drones to obtain their decision}. %(OR Because all players are rational, we can assume that each drone can compute the selected action by each player, if each drone knows the location of all interfering drones and their users).
Each drone finds the direction that maximize the cell spectral efficiency according to equation (\ref{eq:maxutfun_game}).

When all DBSs find themselves in a position that no one wants to change the moving direction, the NE is obtained. Algorithm \ref{alg:gt_alg} is proposed to reach NE at each control mobility time slot. At first, all drones select a random direction from their set of actions. Then each of them finds their best response considering other players' action. Finally, after few trials they all converge to a NE point and move towards the selected directions during the next $t_m$ interval.

\begin{algorithm}
\caption{Game Theory Approach}
\label{alg:gt_alg}
\begin{algorithmic}[1]
\Procedure{}{}
%\State $\text{for each movement time slot}$
\State $NE \gets not\_found$
\For {$each\ p \in \mathcal{P}$}
	\State $rnd \gets random\_number()$
	\State $\theta_p \gets {\mathcal{A}}_p(rnd)$
\EndFor
\While{ $ NE == not\_found$}
% \If{ $trial < trial\_max $ }
% 	\State $trial \gets  trial+1$
 	\For {$each\ p \in \mathcal{P}$}
 		\State $\theta^*_{p} = { arg \  max} \quad u_p(\theta_p, \theta_{-p})$
	\EndFor
	\If{$is\_equal(\theta^*, \theta)$}
	\State $NE \gets found$
	\Else
		\State $\theta \gets  \theta^* $
	\EndIf
%\Else
%\State $exit$
%\EndIf

\EndWhile
\EndProcedure
\end{algorithmic}
\end{algorithm}

%%COMPLEXITY
\subsection{SLR DMA}\label{subsec:slr_alg}

GT is less complex than the optimal DMA, but it still requires communication among the drones. Although inter-drone wireless communication is practically feasible, it does consume resources.
Here, we propose a DMA requiring no communication among drones. Each DBS rather selects its moving direction independently, without any knowledge of other DBS' movement. The DBSs will, however, move in a way that will minimize their interference on other active users in neighbour cells.

%To address the interference issue, the DBS's uses the location of elective users in the interfering cells
%
%The movement of each drone changes the created interference on the active users in the neighbor cells.
%As a result,
%in this strategy each drone not only considers the received signal for its own active users,
%but also attempts to reduce the interference on the other active users in the neighbor cells.
In this model,
each drone knows the location of the active users in the interfering neighbour cells as well.
The integrated interference of drone $n$ on the other active users is referred to as \textit{Leakage} \cite{6571248},
and defined as follows,
\begin{align}
 \begin{split}
L_n=\sum_{j \in N, j\not= n, r_{j,n} \leq \kappa}\big(\sum_{\forall u \in {\mathcal{Q}}_{j}} {S(u,n)}\big).
\end{split}
 \label{eq:leakage}
\end{align}
where $S$ is the received signal strength.

%Each drone calculates the \textit{SLR} (Signal to Leakage Ratio) value for every candidate direction for the active users in the neighbour cells. Such SLR value for each user $u$ of drone $n$ can be formulated as
Then, each drone calculates average the \textit{SLR} (Signal to Leakage Ratio) value for every candidate paths for the active users in the cell. Such SLR value for each user $u$ of drone $n$ can be formulated as
\begin{align}
 \begin{split}
SLR_u= \frac{1}{L_n}S(u,n).
\end{split}
 \label{eq:slr}
\end{align}

Each drone selects the direction that maximizes the average SLR for all active users in the cell.
Such direction is formulated as
\begin{align}
 \begin{split}
 \theta_n =& \text{arg}\,\max\ \frac{\sum_{u \in {\mathcal{Q}}_{n}} SLR_u }{|{\mathcal{Q}}_{n}|} \quad \forall\ n \in \mathcal{N}\\
 s.t. \quad &\theta_n \in  [-\theta_{max}^{v} :g: +\theta_{max}^{v}] \quad
\end{split}
 \label{eq:slr_alg}
\end{align}

%%
%\mahbub{How the 10 data points within $t_m$ is averaged not clear. The use of $t$ in some notations is not clear.}
%
%\azade{text is updated}
%%

\subsection{SNR DMA} \label{sec:snr}

The SLR DMA avoids DBS-to-DBS communication, but requires knowledge of active users in interfering cells. Here we propose a simpler algorithm that neither requires DBS-to-DBS communication, nor does it need to know the location of active users in neighbour cells. We call this algorithm \textit{SNR}, and it is based on the maximization of \textit{SNR} (Signal to Noise Ratio) for active users in each cell.

The \textit{SNR} of an active user $u$ associated to drone $n$ can be defined by
\begin{align}
\begin{split}
SNR^{t}_{u}=\frac{S(u,n)}{N_u},
\end{split}
\label{eq:snr}
\end{align}

%Plugging (\ref{eq:noise}) and (\ref{eq:rcvpower}) into (\ref{eq:snr}), yields
%\begin{align}
%\begin{split}
%SNR^{path}(u,n)=\frac{1}{BN'}\times p_{tx} \times A'_{path} \times (d_{u,n})^{-\gamma_{path}},
%\end{split}
%\label{eq:snr_simple}
%\end{align}
%where $N' = 10^{ \frac{-174+\delta_{ue}}{10}}\times10^{-3}$.

Using only the locations of its own active users, each DBS calculates the average SNR for every active user along the candidate paths, and selects the direction that maximizes the average SNR for all active users (as defined in equation (\ref{eq:snr_alg})).

\begin{align}
 \begin{split}
 \theta_n =& \text{arg}\,\max\  \frac{\sum_{u \in {\mathcal{Q}}_{n}} SNR_u }{|{\mathcal{Q}}_{n}|}\quad \forall\ n \in \mathcal{N}\\
 s.t. \quad &\theta_n \in  [-\theta_{max}^{v} :g: +\theta_{max}^{v}] \quad
\end{split}
 \label{eq:snr_alg}
\end{align}

%%
%\mahbub{How the 10 data points within $t_m$ is averaged not clear. The use of $t$ in some notations is not clear.}
%
%\azade{text is updated}
%%

\subsection{Complexity Comparison of DMAs} \label{subsec:complexity}
We consider two types of complexities, \textit{computational} and \textit{signalling}. Computational complexity refers to the number of combinations to be evaluated to find the optimal direction. Signalling complexity on the other hand refers to the amount of needed signalling among drones and users to obtain the required information for a specific DMA.
For example, in \textit{OPT} algorithm, each drone needs to know the location of other drones ($N$), and all users in the system ($ N.U $). Therefore, $N\times (N+N.U)$ signalling is needed in the whole system. On the other hand, a drone following \textit{SNR} algorithm only needs to know the location of active users in its cell, resulting in a total $N.U$ signalling complexity for the system.

In Table~\ref{tbl:complexity}, the proposed algorithms are sorted based on their computational and signalling complexity. As we can see, finding the optimal solution through exhaustive search is the most complex one requiring the most amount of computation as well as signalling. The \textit{SNR} DMA, on the other hand, has the least complexity.

%\azade{SLR: each drone needs the location of all users ($N.U$), for the whole system it is becoming $N\times (N.U)$.}

\begin{table}
\centering
\caption{Computational and signalling complexities of different DMAs}
\label{tbl:complexity}
\begin{tabular}{|M{0.7cm}|| M{3.6cm} |M{3cm}|}
\hline
\textbf{DMA} &Computational Complexity  &Signalling Complexity \\
\hline
\textbf{OPT} &$O(G^N)$ &  $ O(N.(N+N.U)) $  \\
\hline
\textbf{GT} &$O(N.G)$ &  $ O(N.(N+N.U)) $ \\
\hline
\textbf{SLR} &$O(N.G)$ &  $ O(N^2.U)  $ \\
\hline
\textbf{SNR} &$O(N.G)$ &  $ O(N.U) $  \\
\hline
\end{tabular}
\end{table}

\section{Evaluation} \label{sec:evaluation}

In this section, the performance of our proposed DBS mobility algorithms, as well as the baseline approach where the DBS simply hovers over the centre of the cell, is evaluated using simulations. To be able to use practical values for the drone parameters, such as the flying speed ($v$), maximum acceleration ($a_{max}$), and the minimum possible interval for updating mobility parameters ($t_m$), we conduct some tests with a popular consumer drone called DJI Phantom 4. Before presenting the performance results, we explain the metrics used for the evaluations, the assessment of Phantom 4 parameters, and the simulation setup.
%The numerical results will be compared against that of a fixed drone BS hovering over the center of target area.
%Such baseline scheme is referred to as the hovering (\textit{HOV}) strategy.
% All results have been averaged over 5 runs to smooth out the randomness.

\subsection{Performance Metrics} \label{sec:performancemetric}
Here, we define the required metrics in order to evaluate the system model and its performance.
%All defined metrics are designated for a specific time $t$, however, the subscript $t$ is dropped in this section in order not to make the notations unnecessarily complicated. We will explicitly state otherwise if $t$ should be considered in the corresponding expressions.

\subsubsection{Spectral Efficiency}

The time-averaged SE of the considered $N$-cell system over a given time period $T$ is one of the main metrics used to evaluate the system performance. Equation (\ref{eq:sesystem}) is used to compute system SE ($\bar{\Phi}$) at the start of each \textit{resource allocation slot}. As there are many resource allocation slots in $T$, values of $\bar{\Phi}$ calculated for all slots are averaged to obtain the time-averaged SE.

%To calculate the average spectral efficiently over a given time period, the value of spectral efficiency for all drones at the start of each resource allocation slot is evaluated (as defined in equation (\ref{eq:sesystem})), and is averaged to calculate the average SE.
%The average SE of the considered $N\times N$-cell system is one of the main metrics used to evaluate the system performance, which is defined in equation (\ref{eq:sesystem}).

%\mahbub{I understand that equation (\ref{eq:sesystem}) defines how to compute SE at any given instant of time. It is not clear how you obtain the SE in this section for the entire simulation time.}
%
%\azade{ the SE is calculated at (start of) each resource allocation slot, then  the average is taken for the entire simulation time. the text is updated. the simulation time is not defined yet.}

\subsubsection{Jain Fairness Index}

%According to the definition of the average SE,
%the average user data rate (bits/sec) can be written as
%\begin{align}
% \begin{split}
%\bar{{\mathcal{R}}}_u = \bar{\Phi}_u \times b_u.
%\end{split}
% \label{eq:datarate}
%\end{align}

In the considered multi-user multi-drone system,
we define a fairness metric according to the Jain index to evaluate the fairness among the users,
which is formally presented as \cite{jain1984quantitative}
\begin{align}
 \begin{split}
{\mathcal{J}}(\bar{{\mathcal{R}}}_1,\bar{{\mathcal{R}}}_2, \dots, \bar{{\mathcal{R}}}_U) = \frac{(\sum_{u=1}^{U} \bar{\mathcal{R}}_u)^2}{U\sum_{u=1}^{U} (\bar{\mathcal{R}}_u)^2}.
\end{split}
 \label{eq:jainindex}
\end{align}
where $\mathcal{R}$ is the user data rate (bits/sec).
According to the definition of the average SE,
the average data rate of a user associated with drone $n$ can be written as
\begin{align}
 \begin{split}
\bar{{\mathcal{R}}}_u = \bar{\Phi}(u,n) \times b_u.
\end{split}
 \label{eq:datarate}
\end{align}

\subsubsection{Packet Throughput}
Additionally,
packet throughput,
the ratio of successfully transmitted bits over the time consumed to transmit the said data bits,
can be expressed as
\begin{align}
 \begin{split}
{\mathcal{T}} =  s\times\frac{1}{\tau}
\end{split}
 \label{eq:packet_thp}
\end{align}

Considering all downloaded packets in a cell by all users, the average packet throughout is considered as a performance metric.

Moreover, 
as recommended by the 3GPP~\cite{3gpp36913}, 
the cell edge user throughput is defined as the 5-percentile of CDF of the packet throughput. 
Generally speaking, 
a more homogeneous distribution of the user experience over the coverage area is highly desirable, 
and hence improving the cell edge performance is particularly meaningful in practice.

\subsubsection{Completed Request}
When a data packet finishes transmission,
it is considered as a \textit{completed request}.
In the system,
we measure the number of completed requests for all users.
The average completed requests per user is an \textit{application layer} metrics that can be used to evaluate the system performance.

\subsection{Experimental Assessment of Drone Parameters} \label{subsec:experiments}

In this section, we explain our experiments with Phantom 4 (see Table \ref{tbl:dronespec} for specifications) to obtain the practical ranges for three drone parameters, the flying speed ($v$), maximum acceleration ($a_{max}$), and the minimum possible interval ($t_m$) for updating mobility parameters. The mobility of the Phantom 4 is controlled using an Android application that we developed in-house based on the DJI's software development kit (SDK) \cite{djisdk},\cite{wowmom_ws2017}. In each flight instruction, the velocity in X, Y, and Z direction, height, and flight length can be set.
Flight data, which includes altitude, latitude, longitude, velocity in X, Y and Z direction, battery voltage, and time, are recorded every 100 ms for post-processing.

%In order to validate the drone mobility model in practice, we conducted experiments using a real drone.
%\subsection{Experimental Test-bed} \label{sec:testbed}
%We conducted both real field experiments and simulations using DJI drone \cite{djiPhantom}, namely Phantom 4.

\begin{table}[]
	\centering
    \caption{Phantom 4 mechanical specification \cite{djiPhantom}}
	\label{tbl:dronespec}
	\setlength\arrayrulewidth{1pt}
	\begin{tabular}{|ll|l|}

\hline
		\cellcolor[HTML]{EFEFEF}\textbf{Name} & \cellcolor[HTML]{EFEFEF}Phantom 4 &  \multirow{7}{*}{\includegraphics[height=2.3cm]{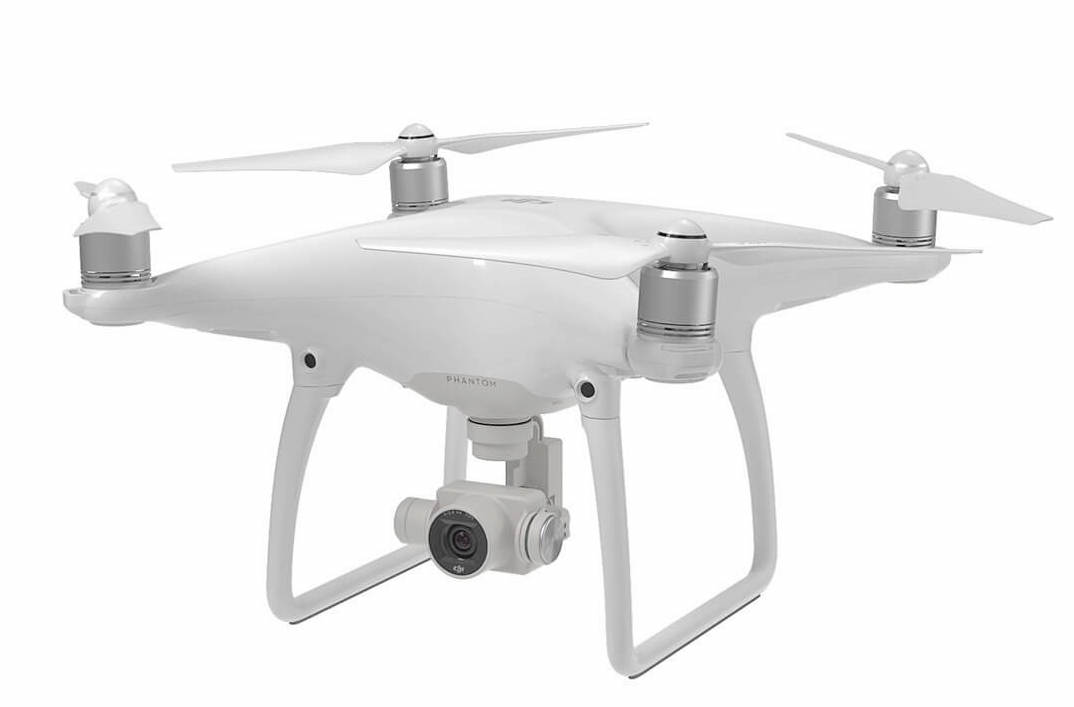}}  \\
		\textbf{Release Date}&March 2016 &                   \\
		\cellcolor[HTML]{EFEFEF}\textbf{Weight}&\cellcolor[HTML]{EFEFEF}1380 g &                   \\
		\textbf{Max Lateral Speed}& 20 m/s   &                   \\
		\cellcolor[HTML]{EFEFEF}\textbf{Max Ascent Speed} &\cellcolor[HTML]{EFEFEF}6 m/s &                   \\
		\textbf{Max Descent Speed}& 4 m/s  &                   \\
		\cellcolor[HTML]{EFEFEF}\textbf{Max Flight Time}&\cellcolor[HTML]{EFEFEF} 28 min &                   \\
		\hline
	\end{tabular}
\end{table}

%Diagonal Size (Propellers Excluded)	350 mm
%\mahbub{show the picture of the drone next to the table (instead of at the top) to save space. }
%
%\azade{updated}

%\subsection{Experimental Results} \label{sec:expr_result}

%We first explore the impact of drone flying speed on its battery life by conducting a set of real flying experiments in an open field.
%Then,
%we use the DJI simulator to study various aspects of drone maneuverability in a windless condition inside our laboratory.
%
%A reasonable height of 10m,
%which is also recommended by the 3GPP as the height for small cell base stations,
%is selected for the drone in all experiments.
%Moreover,
%recent studies shows that the network capacity suffers from a severer degradation as drones fly higher \cite{7842150}.
%Also,
%due to safety reasons and regulation constraints drones can not fly too low.
%

\subsubsection{Drone Flying Speed} \label{sec:batryvsspeed}
The consumer drones can be flown in a 2D plane at a very high speed. For example, Phantom 4 can be flown as high as 20 m/s (see Table \ref{tbl:dronespec}). It is, however, desirable that the continuous mobility of the DBS should not drain the battery faster than the hovering DBS. We therefore need to assess the impact of 2D flying speed on the battery, which will inform our evaluation in terms of the practical drone speeds to consider. For example, it is not useful to evaluate the spectral efficiency gain for a speed that will quickly drain the battery.

%A drone battery life is mostly affected by the mechanical movement of the drone which depends on its speed, mass, and the design \cite{ZORBAS201380,7101619,DiPugliaPugliese2016}.
%Since other hardware elements cannot be controlled,
%it is important to understand how the flying speed affects the battery life.

\begin{figure}
\centering
\includegraphics[ width=0.45\textwidth]{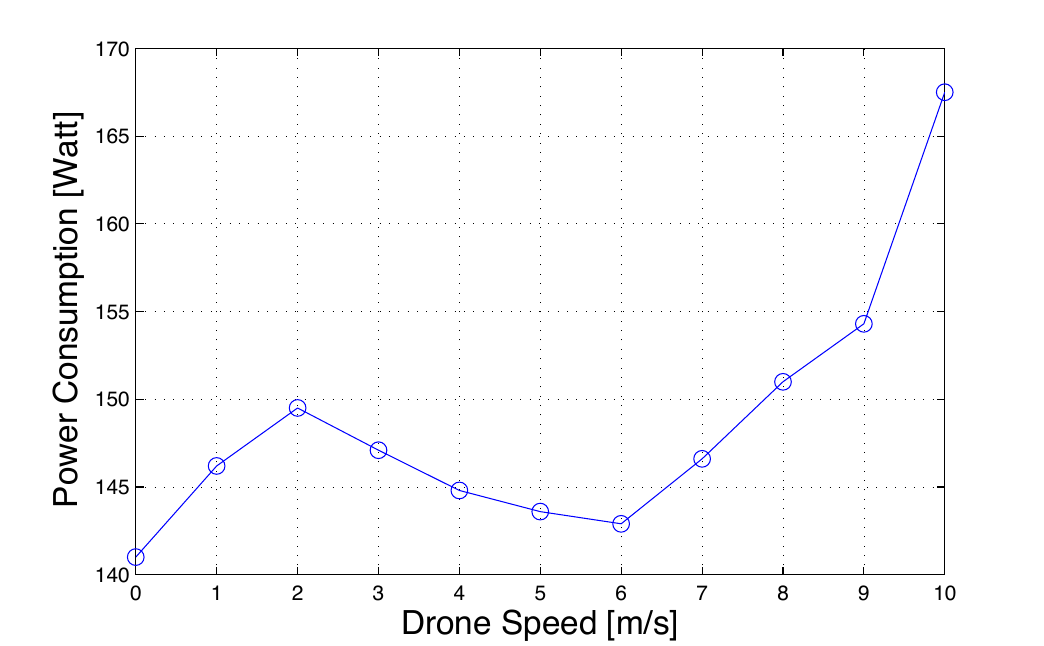}
\caption{Drone power consumption vs. speed at 10 m height}
\label{fig:powervsspeedh10}
%\vspace{-0.4cm}
\end{figure}

To observe the impact of speed on battery life,
we fully charged the battery at the start of each experiment.
Then we flew the drone over an open field in a \textit{way point format},
i.e., between two specified points going back and forward continuously,
until the battery reached 20\%,
which is the minimum the drone can fly on, while keeping the drone altitude at 10 m.
We repeated this experiment for 11 different speeds,
from 0 m/s to 10 m/s with increments of 1 m/s.
For each speed,
we repeated the experiment five times and reported the average power consumption in Figure \ref{fig:powervsspeedh10}.

We observe a very interesting result. The power consumption characteristics below and above the speed of 8 m/s are very different. The amount of power consumed fluctuates below 8 m/s (perhaps due to the wind factor), but it stays below 150 W. On the other hand, power consumption starts to increase rapidly if we fly the drone above 8 m/s.  For example, at 10 m/s, the power consumption is 167 W, which is 11\% higher than that of 8 m/s. In our evaluations, we therefore consider flying speeds up to 8 m/s, which will limit the impact on battery life despite the continuous flying of the drone.

%This means that if Phantom 4 is flown below 8 m/s, the impact of continuous moving will not have any negative effect on the battery compared to a drone hovering over a fixed location.
%
%We observe some fluctuations in power consumption when the speed is less than 8 m/s,
%but our conjecture is that the wind may actually help reducing power consumption sometimes if the drone is flying in the same direction of wind flow (tail wind).
%Unfortunately,
%the DJI did not have a mechanism to record wind data for us to further analyse power consumption fluctuations below 8 m/s.
%Wind analysis remains part of our future study.
%

%The remaining experiments are conducted inside our lab using the DJI simulator.

\subsubsection{DBS Direction Update Interval} \label{sec:maxfreq}

%\mahbub{for DJI simulator, if the drone is not flying, why you need to connect the drone to the laptop?  Why the commands from the android cannot be directly input to the simulator in laptop?}

Ideally, it would be useful for the DBSs to be able to update their directions at an arbitrarily small interval, so they could respond quickly to the dynamics of the system. In practice, however, the value of $t_m$ would be limited by the the drone hardware. To guide our system evaluations, we therefore conduct some experiments with a consumer drone, Phantom 4, to obtain some idea about the practical values for $t_m$.

%We use the Android app for two kinds of experiments.
%The first kinds of experiments are conducted in an open field where we actually fly the drone using the Android app.
%The second kinds of experiments, which we call emulations, are done entirely within the laboratory by DJI Simulator and the Android app.
%DJI provides a flight simulator that allows to connect a propellers-less drone to the simulator,
%which simulates drone flight by taking all control commands from the drone.
%The drone therefore actually does not fly (propellers are taken off), but the simulator can show the drone movements on a map and record all flight data.
%This allows experimenting with a real drone without leaving the laboratory and without being affected by wind.
%We have found that this type of experiment is more reproducible due to lack of unpredictable influence from the weather. %We used DJI Assistant 2 for emulating Phantom 4.

\begin{figure}
\centering
\includegraphics[scale = 0.3]{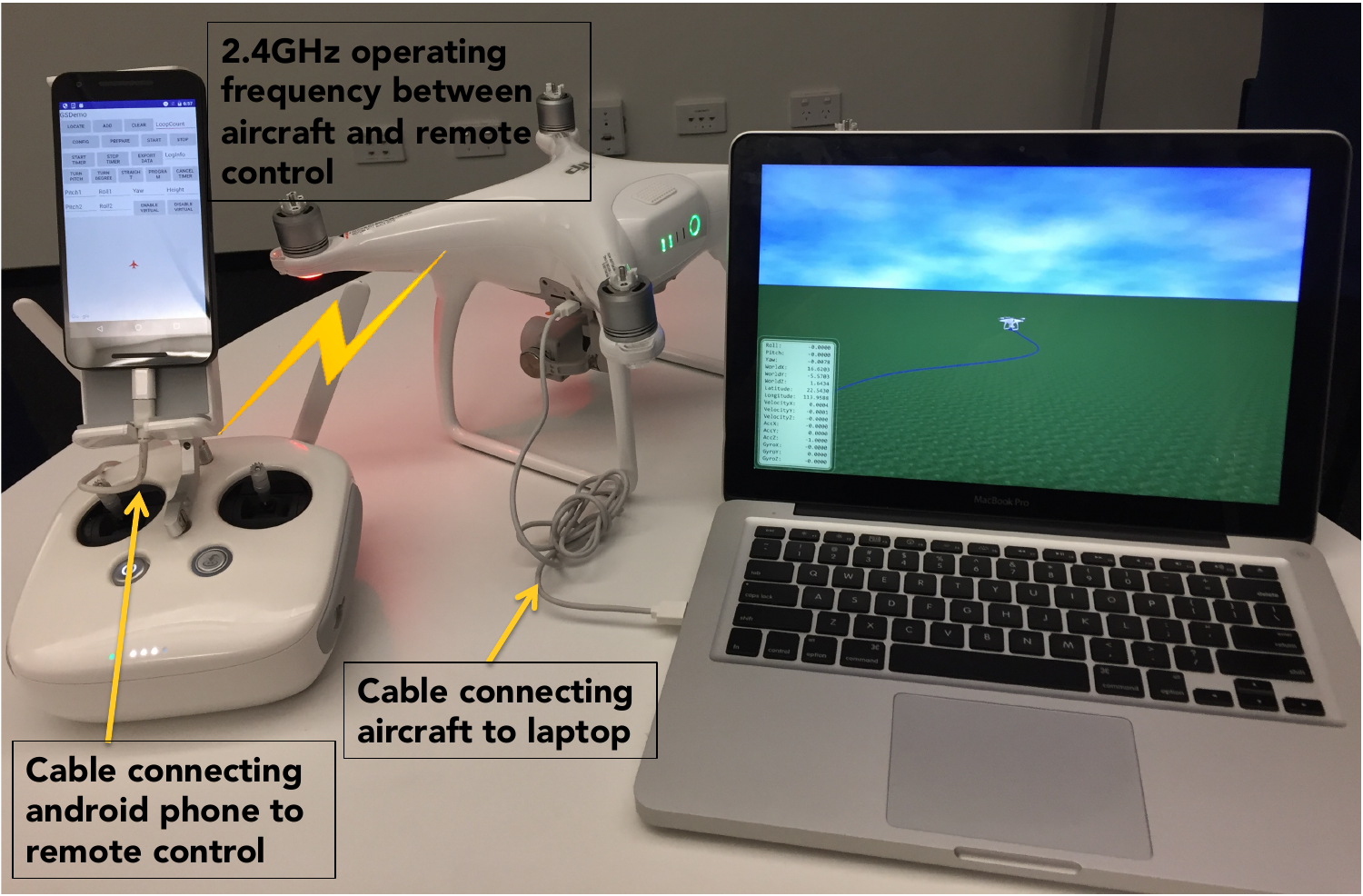}
\caption{Emulated Phantom 4 (propellers-less) in DJI assistant 2 using our developed Android application}
\label{fig:simulator}
%\vspace{-0.4cm}
\end{figure}

We use the flight simulator, DJI Assistant 2, together with an Android application we developed to force the Phantom 4 follow the designed movements and collect flight records.
Figure \ref{fig:simulator} shows the set up of our experimental test-bed where the propellers of the Phantom 4 are taken off, so it actually does not fly, but provides input to the simulator via the connected cable.

%We conducted experiments to figure out what is the maximum frequency at which the drone can be instructed to change its direction (take turns).
%For our application that involves dynamic elements,
%the drone directions have to change frequently.
%The performance of our application can be increased if high frequency turning is possible for the drone.
%

To find out the minimum possible value for $t_m$, we designed an experiment where we change the velocity value of the X direction every $t$ sec,
while keeping Y direction fixed. The velocity in X direction, $v_x$, changes to positive and negative values periodically to simulate a zigzag movement.
Both the velocity in X and velocity in Y direction are selected based on the cruising speed and the target turning angle as presented in the following equations
\begin{align}
 \begin{split}
v_x = \pm v.\sin(\theta/2); \quad v_y = v.\cos(\theta/2)
\end{split}
 \label{eq:velocityxy}
\end{align}
where $v = \sqrt{v_x^2 + v_y^2}$ is the cruising speed,
and $\theta$ is the issued turning angle in radians.
We choose a very low flying speed and a small turn command to avoid hardware restrictions play a part.
To be precise, the drone is moving with a speed of 2 m/s and the turning angle command is 0.1 rad ($\approx$ 5 degree).
We keep the drone height fixed to 10 m, so the $v_z$ is zero. With these settings, the only drone variable that should change due to the turning commands is its $v_x$.

\begin{figure}
\centering
\includegraphics[ width=0.5\textwidth]{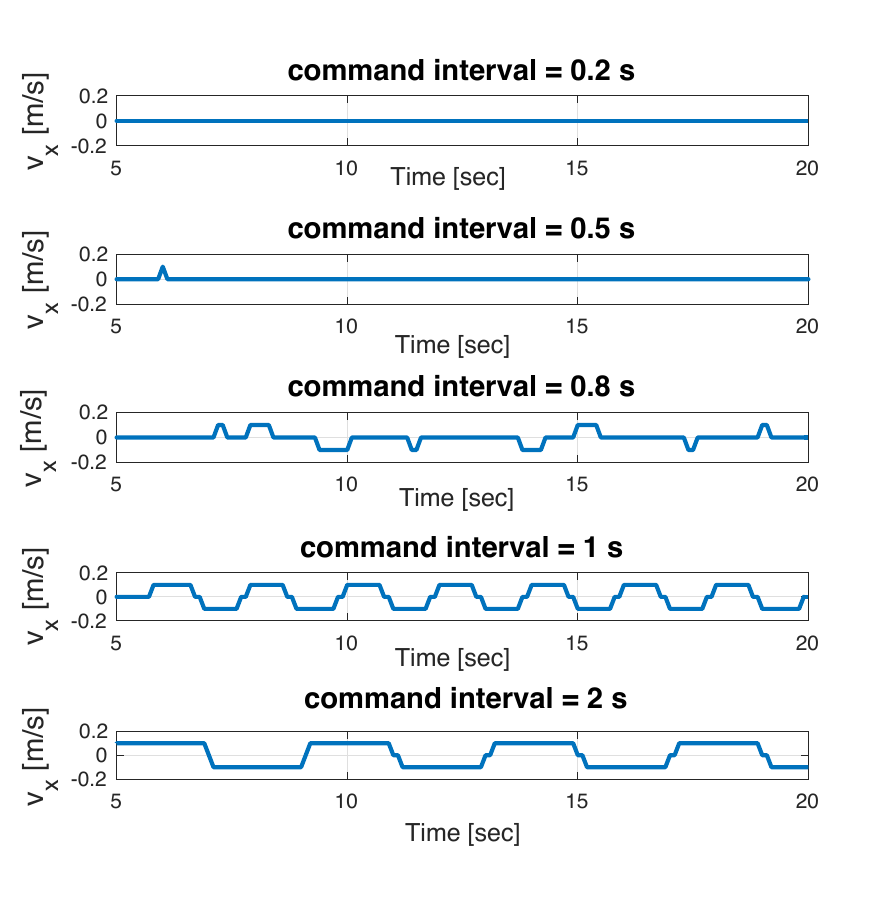}
\caption{Variation of drone $v_x$ with different command intervals}
\label{fig:v_x}
\end{figure}

We recorded $v_x$ every 100 ms and plotted the \textit{change} in Figure \ref{fig:v_x} for five different command intervals ($t$), 0.2 sec to 2 sec. As we can see in Figure \ref{fig:v_x}, the $v_x$ does not change during the flight if the command intervals are less than 1 sec. However, the $v_x$ changes according to the commands when the commands are issued at intervals of 1 sec or higher. This experiment confirms that there is a minimum value of $t_m$ for a given drone hardware and make and it may be around 1 sec.

\subsubsection{Maximum Acceleration}\label{sec:accelaration}

To find out the maximum acceleration that our Phantom 4 can exert while taking turns with a constant speed, we repeated the zigzag experiments with different command values for the turning angle, i.e., the $v_x$ and $v_y$ %commands values
were adjusted to give a turning command with a specific angle.  We used $t_m = 1$, i.e., we commanded the drone to complete the turn within 1 sec. We start the experiments with a small turning angle and increase the value of the turning angle gradually, monitoring both the average speed and the actual angle turned (from the observed $v_x$ and $v_y$, the actual angle turned is obtained
as $2 \times \arctan (\displaystyle \frac{v_x}{v_y})$). From the observed speed and angle,  we obtain the acceleration using equation (\ref{eq:agilityformula}).

\begin{figure}
	\centering
	\includegraphics[ width=0.45\textwidth]{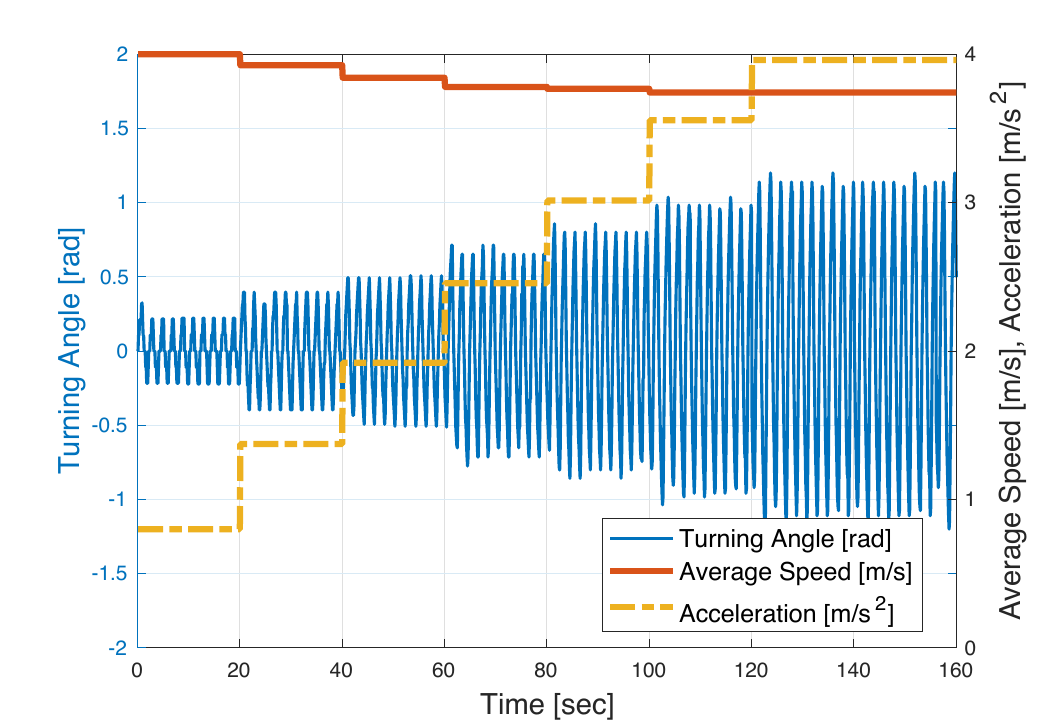}
	\caption{Average drone speed, instantaneous turning angle, and acceleration during different zigzag experiments with various commanded turning angle}
	\label{fig:turn_acc}
\end{figure}

The results are shown in Figure \ref{fig:turn_acc}. We can see that in the beginning, acceleration increases within increasing turning angles while the average speed remains close to the instructed value of 4 $m/s$. The acceleration increased from 1.56 $m/s^2$ to 4 $m/s^2$ in 120 sec., as we increased the turning angle value from 0.25 rad to 1.2 rad. This means that the drone was able to increase its acceleration to meet the increase in the turning demand during this time period. However, after 120 sec., the drone cannot meet the increase in turning demand anymore and its acceleration saturates to approximately 4 $m/s^2$. This experiment clearly shows that drones have a maximum acceleration, which is approximately 4 $m/s^2$ for Phantom 4. We will use this value as a baseline in our simulations.

\subsection{Simulation Setup} \label{subsec:simulation_setup}

We use MATLAB to simulate the proposed DroneCells system with multiple cells and multiple mobile users in each cell. Due to inter-cell interference, \textit{outer} cells in the simulated network scenario will receive less interference than \textit{inner} cells. To obtain unbiased performance results, data is collected only from inner cells. More specifically, we follow the 3GPP approach and create three tiers of neighbour cells around an interested inner cell~\cite{3gpp36814}. A total of 49 square cells are considered in our simulation and data is collected only from the centre cell.

The grid cell size, number of users and their traffic model follow the parameters recommended by the 3GPP~\cite{3gpp36814}, and are shown in Table~\ref{tbl:params}. Our preliminary simulation results show that the system performance becomes stable after 500 seconds. As a result, we run all simulations for 800 seconds to obtain meaningful results.
Moreover,
to mitigate the randomness of the results,
all results have been averaged over 10 independent runs of 800-second simulations.

%5 mobile users are deployed,
%which generate data requests of 4MByte packets where reading time follows an exponential distribution with a mean of 40 seconds.

%The other parameters and their values used in our simulations are listed in Table~\ref{tbl:params}.

\begin{table}[]
\centering
\caption{Definition of parameters and their value}
\label{tbl:params}
\begin{tabular}{lll}
\hline
{\bf Symbol}              & {\bf Definition}    	 & {\bf Value}  	 \\
\hline
\hline
$N$					& Number of Drones	& [9, 49]\\
$B$					& Total Bandwidth   &5 MHz\\
$U$			&Number of Users in Each Cell&	5		\\
$h$		& Drone Height         &10 m\\
$v$		& Drone Speed		&[2, 4, 6, 8] m/s	\\
$w$		& Edge Length of a Square Cell	&80m	\\
$f$		&Working Frequency		&2 GHz		\\
$p_{tx}$& Drone Transmission Power 	 & 24 dBm	\cite{TR36.828}\\
$\lambda$		& Mean Reading Time	& 40 sec \\
$\alpha, \beta$ & Environmental Parameter for Urban Area& 9.61 , 0.16 \cite{yaliniz2016efficient}    \\
$\gamma$	& Path Loss Exponent (LoS/NLoS)& 2.09/3.75 \cite{TR36.828}\\
$\delta_{ue}$     & UE Noise Figure      & 9 dB      \\
${t}_{m}$   &Direction Update Interval 	&1 sec \\
$t_{r}$   &Resource Allocation Slot	&20 msec \\
$\kappa$   &Interference Distance &  200 m\\
$s$		&Data Size	& 40MByte		\\
$G$   & Number of Candidate Directions&21  \\
\hline
\end{tabular}
\end{table}

\subsection{Performance Results} \label{subsec:performance_rslt}

In this section, we evaluate and quantify the potential performance improvements that can be achieved by allowing continuous movement of the DBS in the proposed DroneCells networks. We compare the performance of different DBS mobility algorithms against the baseline scenario where the DBS hovers over the central location of the cell, as well as the case of \textit{optimal} DBS mobility. We analyse performance in terms of spectral efficiency, packet throughput, and request completion rates. We also compare the effect of different resource allocation strategies in  terms of their spectral efficiency and fairness. Finally, we study the benefit of DroneCells under different user densities.

\subsection*{Spectral Efficiency} \label{results:SE}

\begin{figure}
	\centering
	\includegraphics[scale=0.45]{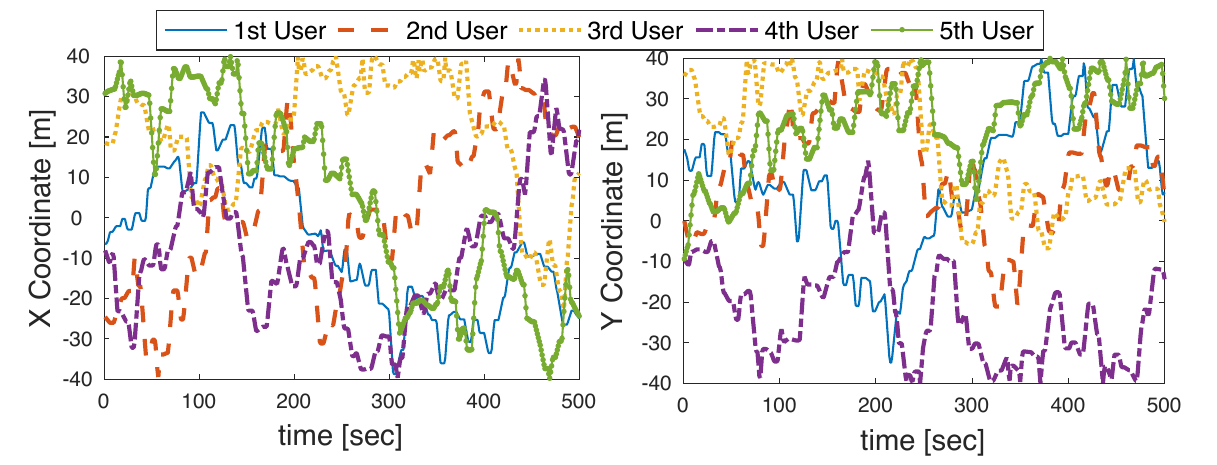}
	\caption{Users mobility pattern}
	\label{fig:userspath}
\end{figure}

The key motivation behind constant movement of the DBS is to ensure that the DBS always move in a way that ultimately reduces the distance between the BS and the user. This is particularly challenging when the users are moving independently in different directions and not clustering together. To demonstrate that the mobile users indeed are not clustering in our simulations, we plot, in Figure~\ref{fig:userspath}, the location coordinates of five users in the centre cell. As we can see, at any given time, different users are located at different places, making the mobility of the DBS a challenging problem.

\begin{figure}
	\centering
	\begin{subfigure}[b]{0.40\textwidth}
		\includegraphics[scale=0.45]{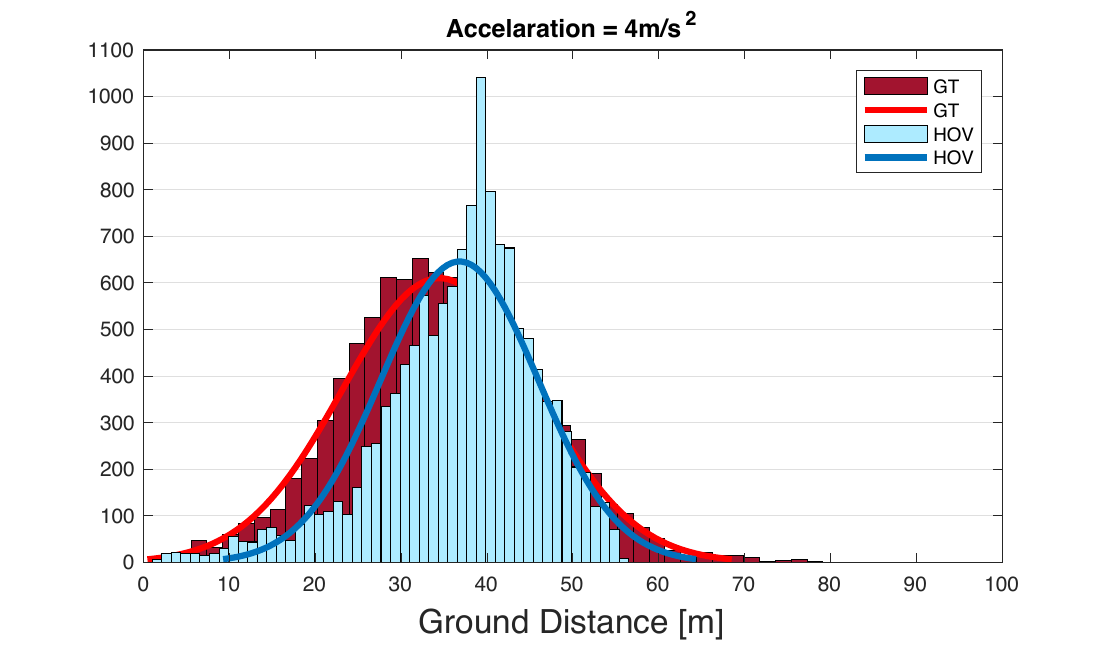}
		\caption{}
		\label{fig:histo_distance}
	\end{subfigure}
	\begin{subfigure}[b]{0.40\textwidth}
		\includegraphics[scale=0.51]{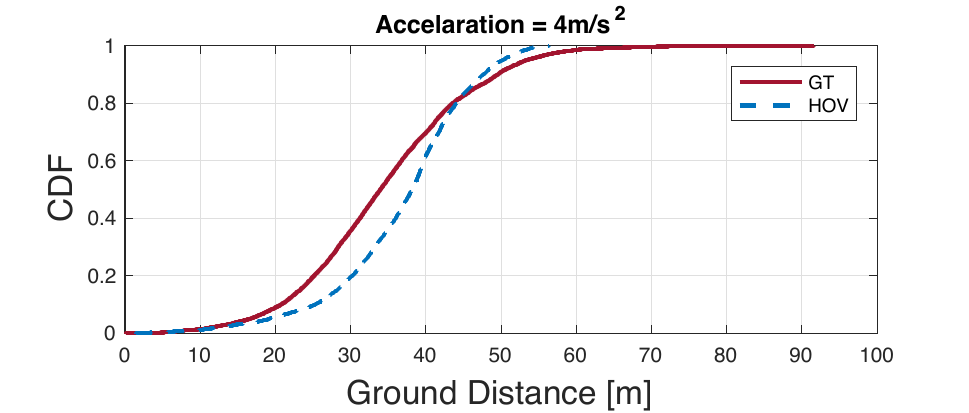}
		\caption{}
		\label{fig:cdf_distance}
	\end{subfigure}
	\caption{(a) Histogram and (b) Empirical CDF of DBS-to-user distance for \textit{GT} and \textit{HOV}}
	\label{fig:grounddistance}
\end{figure}

The problem of reducing the DBS-to-user distance is particularly challenging because, with the freedom to move, the DBS has the potential to actually increase the distance beyond the maximum possible distance of the baseline scenario. For example, with square cells, the maximum possible DBS-to-user distance for DroneCells is the length of the diagonal, which is twice the distance in baseline case.

Despite this challenge, all of the three DBS mobility algorithms were able to reduce the DBS-to-user distance compared to the hovering case. Due to space limits, we use an example from the Game Theory (GT) algorithm to illustrate this outcome. We collected the \textit{ground distance} statistics between any active user and its corresponding DBS during the entire simulation time. Figure \ref{fig:grounddistance} shows the histogram and the empirical CDF of ground distances for the proposed GT algorithm and the baseline model, where drones are moving with the speed of 2m/s for GT and all drones have an acceleration of $4m/s^2$. We can see that the baseline has no data for greater than 56 m, which is the distance from the centre of the square to one of its corner, but the GT has some data points all the way to 93 meter. However, the probabilities for short distances up to 35 m are comparatively very high for GT, which is expected to bring improvements in SNR and ultimately the overall SE of the network.

\begin{figure}
	\centering
	\begin{subfigure}[b]{0.40\textwidth}
		\includegraphics[scale=0.50]{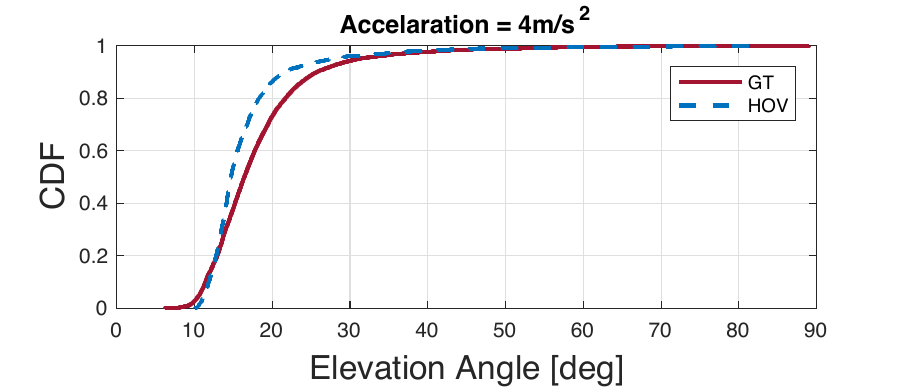}
		\caption{}
		\label{fig:cdf_elevation}
	\end{subfigure}
	\begin{subfigure}[b]{0.40\textwidth}
		\includegraphics[scale=0.50]{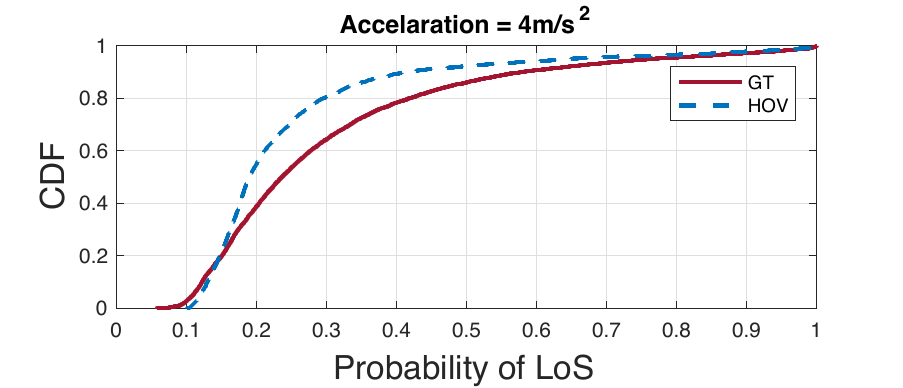}
		\caption{}
		\label{fig:cdf_plos}
	\end{subfigure}
	\caption{Empirical CDF of (a) elevation angle (in degree), and (b) the probability of LoS connection for \textit{GT} and \textit{HOV}}
	\label{fig:elevation_plos}
\end{figure}

Additionally, we collected the elevation angle and accordingly the probability of having LoS connection between active users and their corresponding DBS during the entire simulation time. Figure \ref{fig:elevation_plos} shows the empirical CDF of elevation angle (in degree), and the probability of LoS for the proposed GT algorithm and the baseline model, where drones are moving with the speed of 2m/s for GT and all drones have an acceleration of $4m/s^2$. It can be observed that GT algorithm effectively pushes the elevation angle CDF rightward, resulting in significant improvement in increasing the probability of having LoS connection between active users and drones.

\begin{figure}
	\centering
	\includegraphics[scale=0.45]{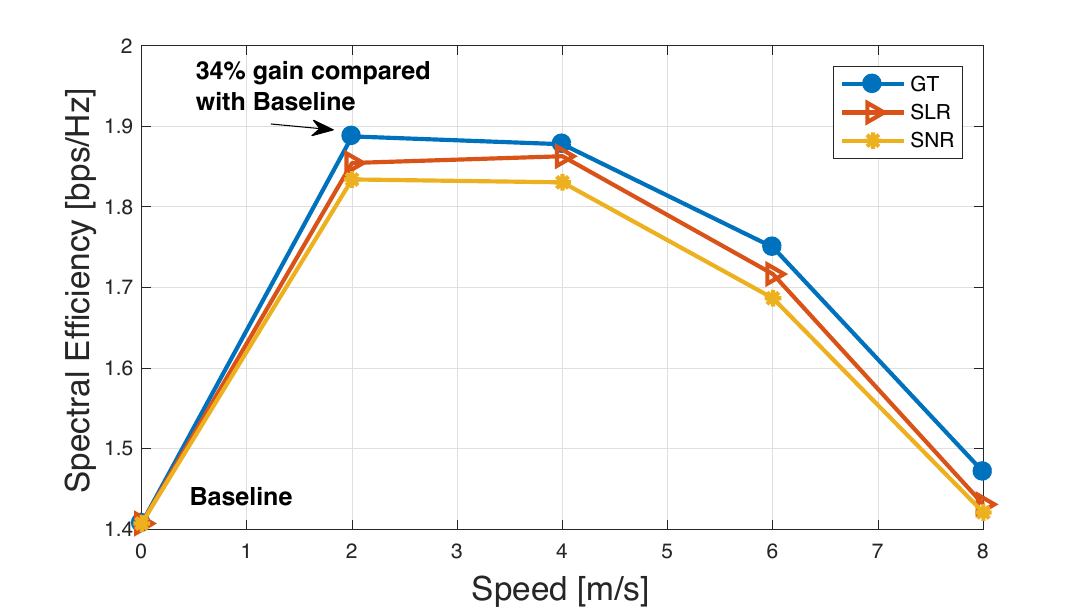}
	\caption{Spectral efficiency of different DMAs for an acceleration of $4m/s^2$}
	\label{fig:se_allalg_c4}
\end{figure}

Figure \ref{fig:se_allalg_c4} shows the spectral efficiency for drones with the acceleration value of $4m/s^2$, where the zero-speed represents
the baseline \textit{HOV} scenario. We can draw the following observations:
\begin{itemize}
  \item Surprisingly, the spectral efficiency does \emph{not} necessarily increase with faster drones for a given acceleration. Instead, there exists an optimal speed to achieve the largest spectral efficiency. This is because, although flying the drone faster may help taking the DBS from one location to another in less amount of time, the higher moving speed reduces the maximum turning angle limiting the possible directions the DBS can move. Using the current Phantom drones, an improvement of up to 34\% in terms of the spectral efficiency can be obtained by moving the drones at a low speed of 2 m/s, which incurs no negative effect on drone energy consumption.
%  \item With a poor choice of drone speeds,
%  e.g., a higher speed than 5m/s,
%  the spectral efficiency of the moving drones could be even less than that of the hovering drones.
  \item The tradeoff between performance gain and complexity is obvious, i.e., the performance of the \textit{GT} strategy is slightly better than that of the \textit{SLR} strategies, which in turn shows some improvement over the \textit{SNR} strategy. Note that the algorithm complexity shows the same order.
\end{itemize}

%The most obvious finding is that drones movement can hugely improve the spectral efficiency compared with the baseline method.
%[height=1.8in, width=0.45\textwidth]

\begin{figure}[]
\centering
\includegraphics[scale = 0.45]{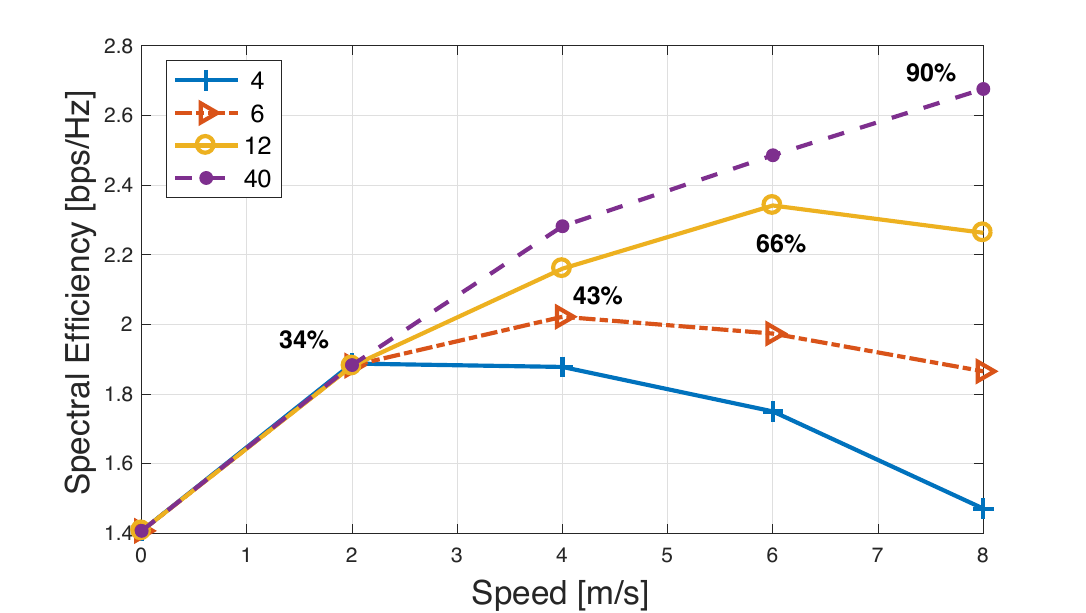}
\caption{Impact of maximum acceleration on spectral efficiency.}
\label{fig:selected_agility_gt}
\end{figure}

To investigate the impact of maximum acceleration on spectral efficiency, we provide more results with various $a_{max}$ values in Figure~\ref{fig:selected_agility_gt} for the \textit{GT} algorithm. Assuming that the acceleration of drones can be improved by factors of 1.5, 3, and 10 in the future, accelerations of 6, 12, and 40$m/s^2$ are considered and compared with the value of $4m/s^2$ (Phantom 4). From Figure~\ref{fig:selected_agility_gt}, we can draw the following observations:
\begin{itemize}
  \item Not surprisingly, increasing the acceleration yields a better spectral efficiency, due to the maneuverability improvement. Choosing the optimal speed as discussed before, %in Subsection~\ref{subsec:SE},
  the spectral efficiency gain ranges from 34\% (2 m/s) to 90\% (8 m/s) as the acceleration increases from $4m/s^2$ to $40m/s^2$,
  respectively.
  \item By increasing the acceleration, the optimal speed for drones to move around is increased as well, as drones are able to enjoy both higher speed and higher manoeuvrability.
  \item If drones have to flown at a very low speed (say, of 2 m/s), increasing the turning angles (higher acceleration) does not have noticeable impact on spectral efficiency. The benefit of higher acceleration can only be reaped by allowing drones to move at a high speed.
 \end{itemize}

%[height=2.2in, width=0.52\textwidth]

The achievable spectral efficiency of different DBS mobility algorithms are summarized in Table \ref{tbl:se_alg} and the gains are compared with the \textit{HOV} baseline model (the SE values are related to the optimal drone speed at each acceleration value). These results show that \textit{GT} outperforms \textit{SLR}, \textit{SNR}, and \textit{HOV} by up to 6\%, 12\%, and 90\%, respectively (for acceleration $40m/s^2$).

%Given more agile drones, \textit{GT} defeats other strategies largely. This is not valid (why difference increases by more agile drone?!)}

\begin{table}[]
\centering
\caption{Spectral efficiency (bps/Hz) and gains for different DMAs at their optimal speeds }
\label{tbl:se_alg}
\begin{tabular}{lllll}
\hline
{}  & {\bf \textit{GT}}    	 & {\bf \textit{SLR}}  	 & {\bf \textit{SNR}}    & {\bf \textit{HOV}}   \\
\hline
Acceleration $4m/s^2$ & 1.88 (34\%)    	 & 1.85 (32\%) 	 & 1.83 (30\%)  & 1.40 \\
\hline
Acceleration  $6m/s^2$  & 2.02 (43\%)  	 & 2 (42\%)	 & 1.94 (40\%)   & 1.40 \\
\hline
Acceleration $12m/s^2$  & 2.34 (66\%)   	 & 2.25 (60\%)	 & 2.18 (55\%)   &1.40 \\
\hline
Acceleration $40m/s^2$  & 2.67 (90\%)   	 & 2.59 (84\%)	 & 2.50 (78\%)   & 1.40 \\
\hline
\end{tabular}
\end{table}

\begin{figure}[]
	\centering
	\includegraphics[scale =0.50]{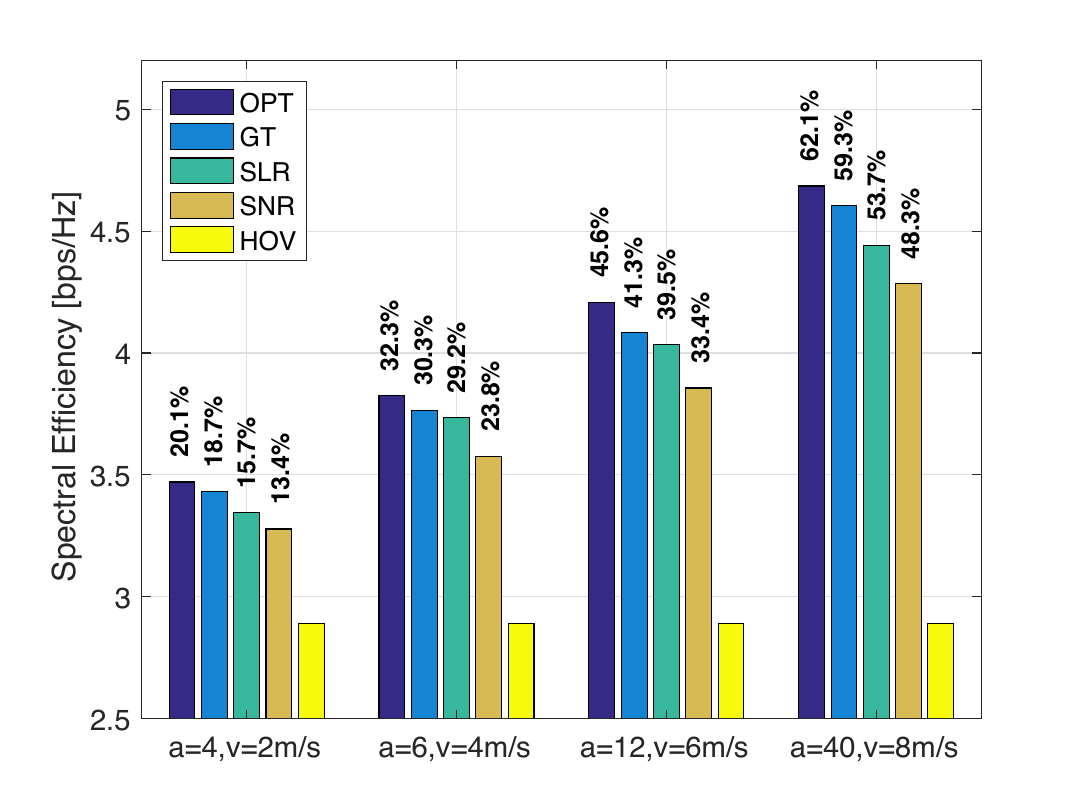}
	\caption{Comparing heuristic results against the optimal DMA.}
	\label{fig:all_opt}
\end{figure}

One important question we have not yet answered is: how good are the proposed heuristics compared to the optimal DMA? Due to the prohibitively high complexity of searching the optimal solution for Problem ~\ref{eq:opt}, we were only able to conduct the exhaustive search for a network scenario of 9 cells with just 1 tier of interfering cells. The results of the optimal mobility control algorithm based on exhaustive search are compared with our heuristic algorithms in Figure~\ref{fig:all_opt} for various accelerations. The key observation is that, for all of the investigated accelerations, only up to 4\%  further improvement can be obtained by the exhaustive search. Considering the extremely high complexity of the optimal DMA, our proposed heuristic algorithms are thus definitely much more useful for practical usage.

\subsection*{Packet Throughput} \label{subsec:celledgeusers}

\begin{figure}
    \centering
    \begin{subfigure}[b]{0.40\textwidth}
        \includegraphics[scale=0.45]{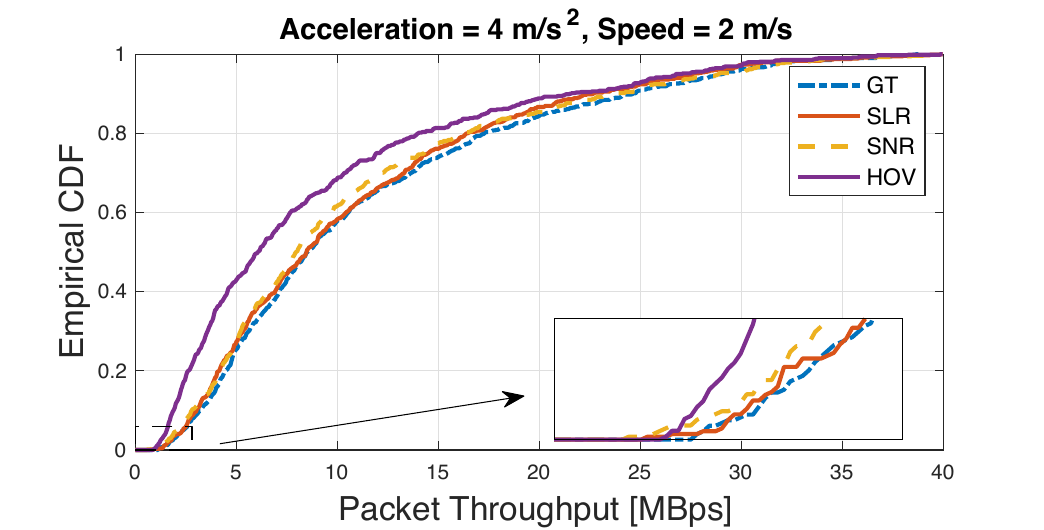}
        \caption{}
        \label{fig:cdf-c4}
    \end{subfigure}
    ~ %add desired spacing between images, e. g. ~, \quad, \qquad, \hfill etc.
    \begin{subfigure}[b]{0.40\textwidth}
        \includegraphics[scale=0.45]{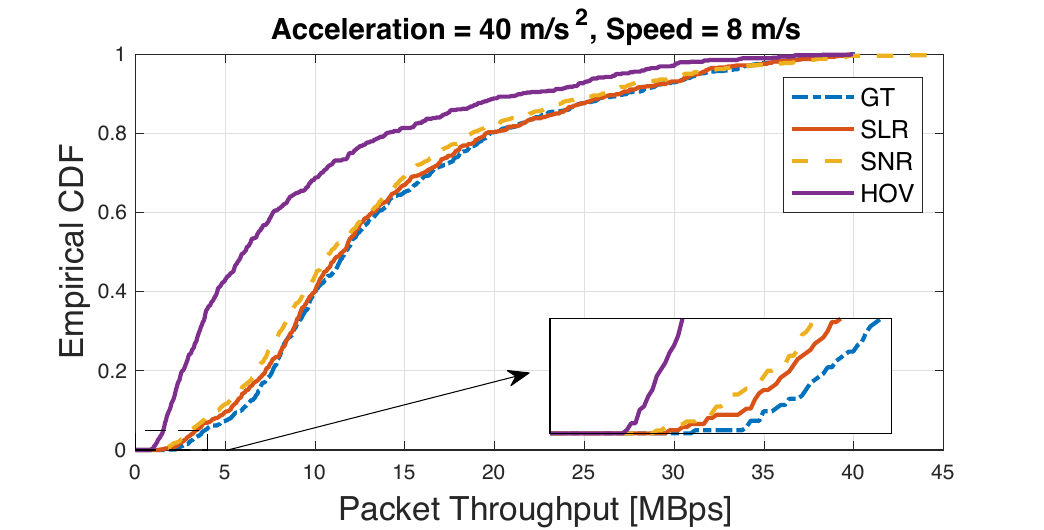}
        \caption{}
        \label{fig:cdf-c40}
    \end{subfigure}
    ~ %add desired spacing between images, e. g. ~, \quad, \qquad, \hfill etc.
    \caption{Empirical CDF for packet throughput with (a) acceleration = $4m/s^2$, and (b) acceleration =  $40m/s^2$}\label{fig:cdf}
\end{figure}

\begin{figure}
\centering
\includegraphics[scale=0.50]{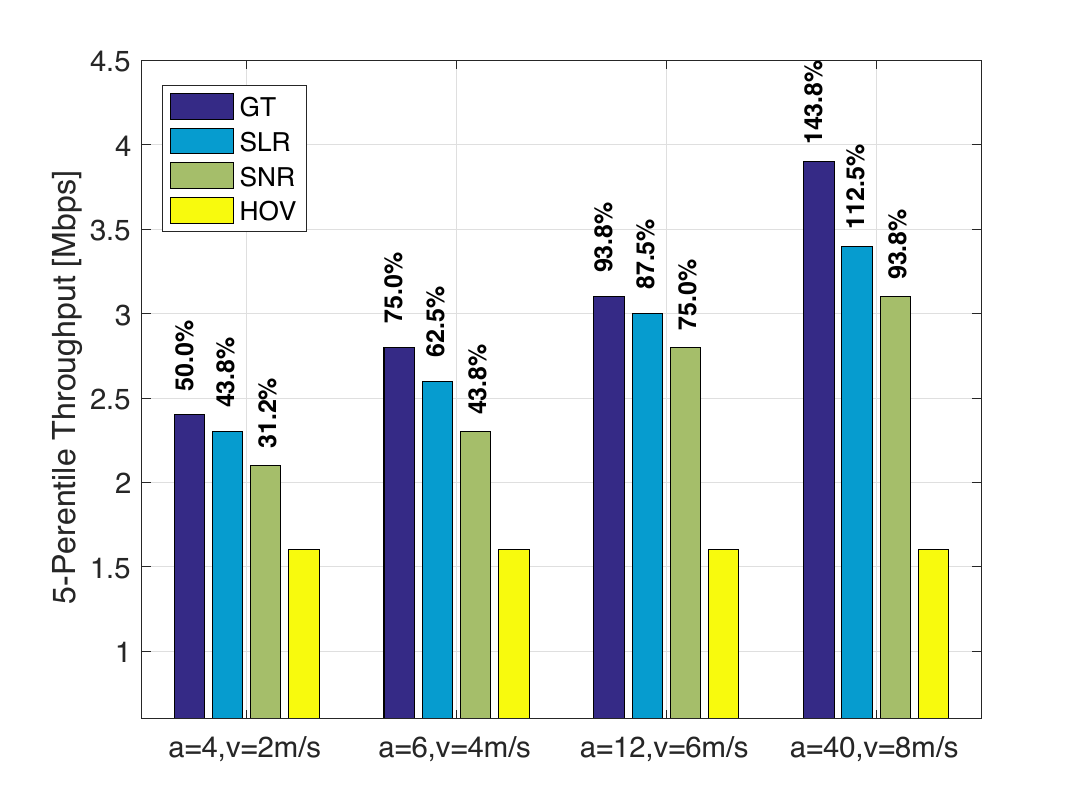}
\caption{5-percentile throughput for different accelerations}
\label{fig:5_percentile}
\end{figure}

To show the performance of the packet throughput,
in Figure \ref{fig:cdf},
we plot the empirical CDF of the packet throughput for the acceleration of $4m/s^2$and $40m/s^2$, when drone's speed is 2 m/s and 8 m/s, respectively.
Moreover,
to quantify
%the average packet throughput and
the 5-percentile packet throughput,
we show these results of the investigated algorithms in Figure~\ref{fig:5_percentile} .
% and Figure~\ref{fig:5per_pkt_thp}.

From Figures~\ref{fig:cdf} and~\ref{fig:5_percentile},
we can draw the following observations:
\begin{itemize}
  \item Compared with the baseline hovering strategy,
  our proposed algorithms successfully push the packet throughput CDF rightward,
  showing significant gains in terms of this performance metric.
  \item
  There is a large performance gain in terms of the 5-percentile packet throughput,
  reaching up to 50\% and 143\% improvement with the existing consumer drones (an acceleration of $4m/s^2$) and the future drones (an acceleration of $40m/s^2$), respectively.
  This is because our algorithms allow drones to move to the vicinity of users,
  while hovering drones are stationary at the cell centre and thus cannot deliver satisfactory QoS to cell-edge users.
\end{itemize}

\subsection*{Request Completions}

Table~\ref{tbl:requestno} presents the average number of completed requests for \textit{GT} algorithm and compares it with that of the \textit{HOV} baseline. As we can see, by increasing the acceleration, the drone can serve a larger number of requests generated by users. On average, the number of completed requests is increasing by 7.2\% (from 84.7 to 90.0) and 14.04\% (from 84.7 to 96.6) with the existing consumer drones (an acceleration of $4m/s^2$) and the future drones (an acceleration of $40m/s^2$), respectively. This improvement in request completion rates is a natural result of the large spectral efficiency gain shown in \textbf{Spectral Efficiency} subsection. %Subsection~\ref{results:SE}.
\begin{table}[]
\centering
\caption{ Average number of completed requests}
\label{tbl:requestno}
\begin{tabular}{|l||c|c|}
\hline
  {} & \textit{\textbf{GT}} & \textit{\textbf{HOV}} \\
  \hline
Acceleration= $4m/s^2$, Speed = 2 m/s & 90.8 (7.2\%)                 & 84.7                  \\
\hline
Acceleration= $6m/s^2$, Speed = 4 m/s& 91.4 (7.9\%)               & 84.7                    \\
\hline
Acceleration= $12m/s^2$, Speed = 6 m/s & 93.0 (9.7\%)                 & 84.7       \\
\hline
Acceleration= $40m/s^2$, Speed = 8 m/s & 96.6 (14.04\%)                 & 84.7       \\
\hline
\end{tabular}
\end{table}

\subsection*{Impact of Resource Allocation Strategies}

\begin{figure}[]
\centering
\includegraphics[scale = 0.5]{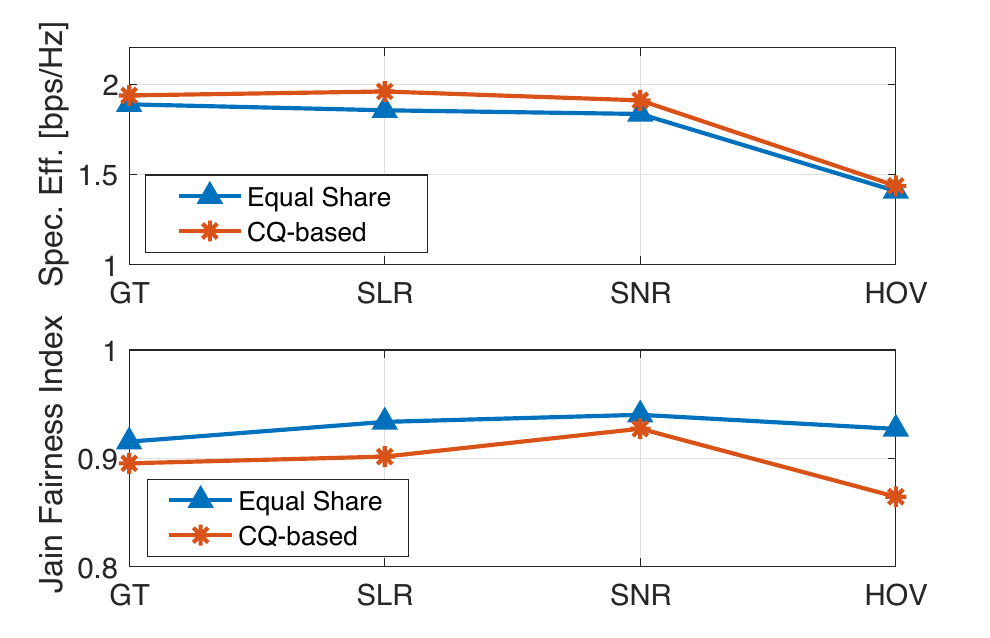}
\caption{Impact of resource allocation strategies on SE and fairness.}
\label{fig:bw_jain}
\end{figure}

Figure~\ref{fig:bw_jain} compares the SE and fairness of the Equal Share and CQ-based resource allocation approaches when the acceleration is $4m/s^2$ and drone speed is 2 m/s. We can clearly see the tradeoff between SE and fairness, i.e., the CQ-based allocation was able to improve the SE at the expense of being less fair. The interesting observation, however, is that the fairness disparity for the CQ-based allocation is only \textit{marginal} for the DroneCells (for all three mobility algorithms), but \textit{significant} for the baseline scenario.

\subsection*{Impact of DBS Direction Update Interval}

\begin{figure*}
	\centering
	\includegraphics[scale=0.48]{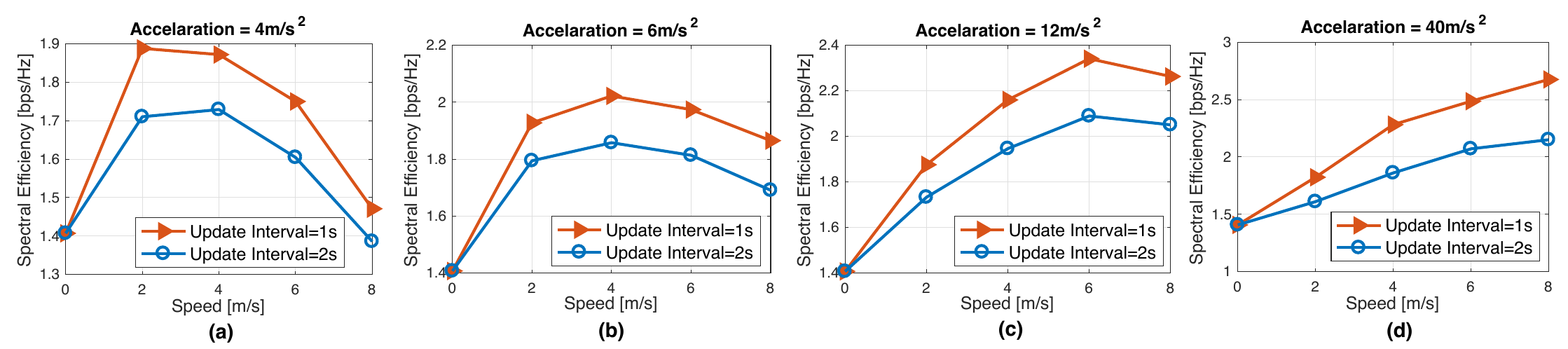}
	\caption{Impact of $t_m$ on spectral efficiency with drone acceleration of (a) $4m/s^2$ (b) $6m/s^2$ (c) $12m/s^2$ (d) $40m/s^2$}
	\label{fig:granularity}
\end{figure*}

%Regardless of the practical limitations of the current drones, w
We have also analysed the impact of direction update interval ($t_m$) on our proposed algorithms. Intuitively, the shorter this interval, the more opportunity the algorithms have to adjust the direction of the DBS and hence are expected to produce better results. However, because we found that 1-sec is the minimum time needed for Phantom 4 to make adjustments, here we consider only $t_m \ge 1$. For various accelerations, Figure \ref{fig:granularity} compares the performance of GT DMA obtained with $t_m=1$ against that of $t_m=2$. As expected, we can see that $t_m=1$ outperforms $t_m=2$ in all cases. More importantly, it is clear that the proposed GT DMA can still improve SE significantly even with a 2s direction update interval. This result confirms that our proposed DroneCells idea has benefit even for low-end drones that may only be controlled with coarser granularity.

\subsection*{Impact of User Density}

\begin{figure}[]
	\centering
	\includegraphics[scale = 0.5]{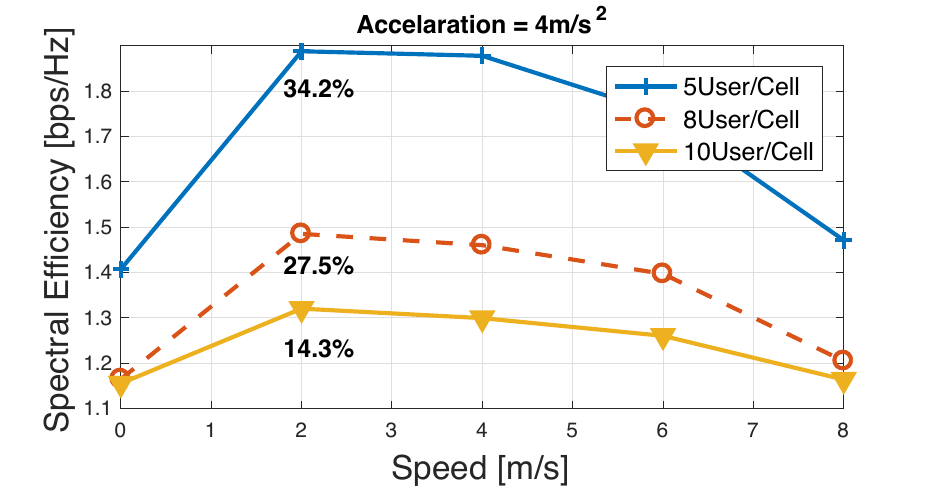}
	\caption{Impact of user density on spectral efficiency}
	\label{fig:userdensity}
\end{figure}

We also explored the impact of user density on spectral efficiency. To this end, the simulation is conducted for higher user density such as 8 and 10 users per cell, and compared with 5 users per cell. Figure \ref{fig:userdensity} presents the average spectral efficiency for \textit{GT} algorithm and compares it with that of the \textit{HOV} baseline.
According to this figure, 
we can observe the followings:
\begin{itemize}
	\item By increasing the user density, the average SE decreases. It can be concluded that the probability of having active users at any time increases by having higher density, resulting in more transmissions for each drone, and higher interference in the system. Indeed, we found that the average transmission times for drones increased noticeably with increasing user density (see Table \ref{tbl:trasnmist_time}).
	\item \textit{GT} improves the spectral efficiency significantly for all user densities illustrated in Figure \ref{fig:userdensity}.
	
\end{itemize}

\begin{table}[]
	\centering
	\caption{Average percentage of transmission time for drones during the simulation time}
	\label{tbl:trasnmist_time}
	\begin{tabular}{|c|c|c|c|}
		\hline
		Density & 5users$ / $cell & 8users$ / $cell & 10users$ / $cell\\
		\hline
		Transmission Time [\%] &51.8\%&78.4\%&91.7\%\\
		\hline
	\end{tabular}
\end{table}

\subsection*{DBS Movements Outside the Cells}
Finally, we evaluate the ability of the proposed DMAs to keep the DBSs within the cell boundary. For different combinations of accelerations and speeds, Table~\ref{tbl:go_outside} presents the percentage of the simulation time (800 sec) the DBSs spend outside their designated cells on average (only GT DMA is shown for space constraint). We can see that the percentage is very small. DBSs would spend less than 1\% of their time outside the cells for speeds less than 6m/s if Phantom 4 (acceleration = 4$m/s^2$) is used. For more agile drones with higher accelerations, the percentage remains below 1\% even for 8m/s.

%
%It is very important to note that in our algorithms there is no guarantee that drones will always stay within their border.
%%Although they have intention to stay within their cells and served their users.
%In this section,
%we calculated the average time that a drone spends outside its border during the simulation time (800sec).
%The results are for two different acceleration cases, i.e.,
%4 and 40 $ m/s^2 $, which are listed in Table~\ref{tbl:go_outside},
%when drones move at various speeds.
%

% \usepackage{multirow}
\begin{table}[]
	\centering
	\caption{Average time a drone spends outside its border during simulation time}
	\label{tbl:go_outside}
	\begin{tabular}{|c|c|c|c|c|}
		\hline
		\multirow{2}{*}{\textbf{Acc.}} & \multicolumn{4}{c|}{\textbf{Speed }}         \\ \cline{2-5}
		& \textit{2$ m/s $} & \textit{4$ m/s $} & \textit{6$ m/s $} & \textit{8$ m/s $} \\ \hline
		\textit{4$ m/s^2 $}& 0s (0\%)    &  0.3s (0.03\%)  & 13.2s (1.6\% ) & 79.5s (9.9\%)\\ \hline
		\textit{40$ m/s^2 $}& 0s (0\%)    &  0s (0\%)  &  0s (0\%) & 0.3s (0.03\%)\\ \hline
	\end{tabular}
\end{table}

\section{Conclusion} \label{sec:conclusion}
In this paper, we proposed mobility control algorithms for drone base stations, which are constantly moving at a fixed height above their cells, in order to improve the spectral efficiency of the system. Extensive experiments and simulations with real drone are conducted to resolve the practical limitation of drones such as their power consumption and manoeuvrability. Applying the practical constraints, it was shown that our proposed algorithms significantly improve spectral efficiency, and packet throughput compared with the hovering drone base stations. These advancements can be brought by low complex algorithms while keeping the drones' energy consumption at the same level as the network where drones are hovering above pre-determined positions.

% or


%\appendices
%% you can choose not to have a title for an appendix
%% if you want by leaving the argument blank
%\section{}
%% use section* for acknowledgment
%\ifCLASSOPTIONcompsoc
%  % The Computer Society usually uses the plural form
%\else
%  % regular IEEE prefers the singular form
%\fi

\ifCLASSOPTIONcaptionsoff
  \newpage
\fi



%\bibliographystyle{IEEEtran}
\section*{Acknowledgment}
The authors gratefully acknowledge Jay Gurnani's contributions in developing the Android application. Azade's research is supported by Australian Government Research Training Program Scholarship and Data61|CSIRO PhD top-up scholarship.

\bibliographystyle{IEEEtran}
\bibliography{tmcarxive}

\end{document}